\documentclass[superscriptaddress,twocolumn,showpacs,preprintnumbers,amsmath,amssymb,floatfix]{revtex4-1}
\usepackage{graphicx}
\usepackage{slashed}
\usepackage{color}
\begin{document}

\title{Extracting the pole  and Breit-Wigner
  properties of nucleon and $\Delta$ resonances from 
  the $\gamma N\to K\Sigma$ photoproduction}

\author{S. Clymton}
\affiliation{Departemen Fisika, FMIPA, Universitas Indonesia, Depok 16424, Indonesia}
\affiliation{Department of Physics, Inha University, Incheon 22212, Republic of Korea}
\author{T. Mart}
\affiliation{Departemen Fisika, FMIPA, Universitas Indonesia, Depok 16424, Indonesia}
\email{terry.mart@sci.ui.ac.id}

\begin{abstract}
  We have developed a covariant isobar model to phenomenologically explain the four possible isospin 
  channels of $K\Sigma$ photoproduction. To obtain a consistent reaction amplitude,
  which is free from the lower-spin background problem, we used the consistent
  electromagnetic and hadronic interactions described in our previous report.
  We used all available experimental data, including the recent MAMI A2 
  2018 data obtained for the $K^0\Sigma^+$ and $K^0\Sigma^0$ isospin channels. 
  By fitting the calculated observables to these data we extract the resonance 
  properties, including their  Breit-Wigner and pole  parameters. Comparison of
  the extracted parameters with those listed by the Particle Data Group yields 
  a nice agreement. An extensive comparison of the calculated observables with
  experimental data is also presented. By using the model we investigated the
  effects of three different form factors used in the hadronic vertex of each 
  diagram. A brief discussion on the form factors is given.
\end{abstract}


\maketitle
\section{INTRODUCTION}
For more than 50 years kaon photoproduction off a nucleon has gained a special interest 
in the hadronic physics community. Early theoretical work was reported in
1957 \cite{masasaki}, but a more comprehensive analysis with fitting to
experimental data was just started in 1966 \cite{thom}. Since then many
efforts have been devoted to explain this reaction, ranging from
quark to hadronic coupled-channel models, as briefly mentioned in the
introduction part of our previous report \cite{brief_intro}.

In the beginning, the main motivations to study kaon photoproduction off 
the nucleon were merely to obtain theoretical explanation of the reaction 
process. However, it has been soon realized that an accurate theoretical model 
describing this elementary process is also useful in many branches of hadronic and 
nuclear physics. In hadronic physics the model is indispensable in the
investigation of the missing resonances that have considerably large
branching ratio to the strangeness channel \cite{missing-d13}, 
the narrow resonance which is also predicted to have a large branching ratio
to the $K\Lambda$ channel \cite{mart-narrow}, and the resonance hadronic coupling 
constant that measures the strength of the interaction between kaon-hyperon 
final state and the resonance \cite{Mart:2013ida}. 
In the nuclear physics this elementary model is required in calculating 
the cross section of hypernuclear photoproduction \cite{mart_hyp}, which is
the main observable in the investigation of hypernuclear spectroscopy.

Recently, the $nn\Lambda$
electroproduction on a tritium target, i.e, the $^3{\rm H}(e,e'K^+)nn\Lambda$
process, has been performed at Jefferson Lab Hall A (JLab E12-17-003)
and the result is currently being analyzed \cite{gogami}. There have been
intensive discussions on whether the $nn\Lambda$ system could be bound
or would lead to a resonant state. Thus, this experiment is expected to shed
light on the $nn\Lambda$ puzzle. Furthermore, an accurate measurement of
hypertriton electroproduction has been proposed and conditionally approved
as the JLab C12-19-002 experiment \cite{gogami}. The experiment is expected
to further elucidate the $\Lambda$ binding energy in the hypertriton, since 
the result of previous measurement indicated a stronger binding energy
\cite{star}. Therefore, an accurate elementary model, describing the 
photo- and electroproduction of kaon on the nucleon target is timely and 
urgently required.

Our previous model used to this end was Kaon-Maid \cite{maid}, which includes
kaon photo- and electroproduction off the nucleon in six isospin channels.
However, due to tremendous increase in the number of 
experimental data, especially in the case of photoproduction, 
Kaon-Maid started to show its deficiency. To get rid of this problem 
we have started to improve the model in the photoproduction sector, 
for which experimental data dominate our present database.

In the previous works we have developed a new and modern covariant isobar 
model for kaon photoproduction off the nucleon 
$\gamma + p \rightarrow K^{+} + \Lambda$ and
$\gamma + n \rightarrow K^{0} + \Lambda$ \cite{previous-work}. The model 
fits nearly 9000 experimental data points and 
employs the consistent hadronic and electromagnetic interactions that 
eliminate contributions of high-spin resonance background 
\cite{Vrancx:2011qv,Clymton:2017nvp}. The latter is widely known as an intrinsic
problem that plagues the formalism of high-spin propagators used to describe
the contribution of nucleon and delta resonances. 
In this paper
we extend the model, based on the covariant effective Lagrangian method, to
include the other four isospin channels in the $K\Sigma$ photoproduction.

We have organized this paper as follows. In Sec.~\ref{sec:formalism} we 
discuss the formalism used in our model. In principle, we use the same 
interaction Lagrangians as described in our previous paper for
the $K\Lambda$ photoproduction, $\gamma + p \rightarrow K^{+} + \Lambda$ 
\cite{Clymton:2017nvp}. In addition, we also briefly discuss the formalism used 
to extract resonance masses and widths at their pole positions in this section.
In Sec.~\ref{sec:result} we present the result of our analysis and compare 
the calculated observables with the available experimental data. To describe
the accuracy of our model in details, an extensive comparison of polarization
observables is given in this section.
In Sec.~\ref{sec:summary} we summarize our analysis and 
conclude the important findings. The extracted Breit-Wigner masses and widths
of the resonances used in the model are listed in Table~\ref{tab:mwbw}
of Appendix~\ref{app:mwbw}.

\section{FORMALISM}
\label{sec:formalism}
\subsection{The Model}
In the present work we consider the photoproduction process of $K\Sigma$ on a nucleon, i.e.,
\begin{eqnarray}
  \label{eq:_photoproduction_process}
  \gamma (k) + N (p) \to K(q) + \Sigma (p_\Sigma) ~.
\end{eqnarray}
Based on the isospin and strangeness conservations, Eq.~(\ref{eq:_photoproduction_process}) implies
four different photoproduction processes given in Table \ref{tab:threshold}.

\begin{table}[!ht]
  \centering
\caption{Four possible isospin channels for $K\Sigma$ photoproduction 
  off the nucleon. The corresponding threshold energies are also listed
  in terms of the photon laboratory energy 
  $k_{0,{\rm lab}}^{\rm thr}$ and total c.m. energy $W^{\rm thr}$.
  }
  \label{tab:threshold}
  \begin{ruledtabular}
  \begin{tabular}[c]{clccc}
    No.&Channel & $k_{0,{\rm lab}}^{\rm thr}$ (MeV)&~~~~~~&$W^{\rm thr}$ (MeV)\\
    \hline
    1& $ \gamma + p$ $\longrightarrow$  $K^{+} + \Sigma^{0}$ &  1046 && 1686\\
    2& $ \gamma + p$ $\longrightarrow$  $K^{0} + \Sigma^{+}$ &  1048 && 1687\\
    3& $ \gamma + n$ $\longrightarrow$  $K^{+} + \Sigma^{-}$ & 1052  && 1691\\
    4& $ \gamma + n$ $\longrightarrow$  $K^{0} + \Sigma^{0}$ & 1051  && 1690\\
  \end{tabular}
  \end{ruledtabular}
\end{table}

\begin{figure*}
\includegraphics[scale=0.48]{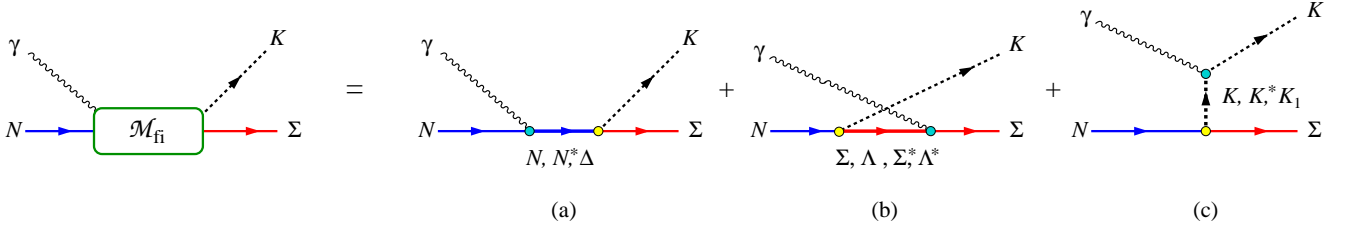}
  \caption{Feynman diagrams for the background and resonance terms of the 
    $K\Sigma$ photoproduction off a nucleon $\gamma(k) + N(p) \to K(q) + \Sigma(p_\Sigma)$.
    The relevant diagrams are grouped according to their intermediate states, i.e., 
    (a) the $s$-channel nucleon, nucleon resonances and $\Delta$ resonances, (b) the 
    $u$-channel $\Sigma$, $\Lambda$ and hyperon resonances, and (c) the $t$-channel
    kaon and kaon resonances.}
  \label{fig:diagrams}
\end{figure*}

The scattering amplitude of these reactions is calculated from the first-order Feynman diagrams 
shown in Fig.~\ref{fig:diagrams}. According to their intermediate states the diagrams can be grouped
into three main channels, i.e., the $s$-, $t$- and $u$-channel, with the corresponding Mandelstam 
variables are defined as
\begin{equation}
s = (p + k)^2 \; ; \;\; t = (q - k)^2 \; ; \;\; u = (p_\Sigma - k)^2\;. 
\label{eq:Mandelstam}
\end{equation}
Note that the notation of the momenta written in Eq.~(\ref{eq:Mandelstam}) is given explicitly in
 Eq.~(\ref{eq:_photoproduction_process}). The corresponding vertex factors can be obtained from the effective Lagrangian 
approach, specifically by using the prescription given in 
Refs.~\cite{Pascalutsa-PRD1998,Pascalutsa-PRC1999,Pascalutsa:2000kd}. 
In our previous study of the $K^+\Lambda$ photoproduction~\cite{Clymton:2017nvp} the Lagrangians were constructed 
according to the method proposed in Ref.~\cite{Vrancx:2011qv} in order to be consistent with the 
formulation of high spin ($J>3/2$) propagators. In the case of $K\Sigma$ photoproduction, the hadronic 
and electromagnetic Lagrangians of the $s$-channel spin-$(n+1/2)$ particle with positive parity reads
\begin{eqnarray}
\label{eq:L_hadronic}
\mathcal{L}_\mathrm{had}&=&\frac{g_{K\Sigma N^*}}{M^{2n+1}}\, \epsilon^{\mu\nu_n\alpha\beta}\,\partial^{\nu_1}\cdots
\partial^{\nu_{n-1}}\bar{\Psi} \, \partial_{\beta} \phi^{*}\, (\gamma_5)^n \, \gamma_\alpha \, \nonumber\\
& & \partial_{\mu} \,\Psi_{\nu_1\cdots\nu_n}\, + \mathrm{H.c.}~, \\
\mathcal{L}_\mathrm{em} &=& \frac{e}{M^{2n+1}}\bar{\Psi}^{\beta_1\cdots\beta_n}\,(\gamma_5)^n\bigl\{g_1\,\gamma_5\,
\epsilon_{\mu\nu\alpha\beta_n} \partial^\alpha \Psi  \nonumber \\
& & + g_2 \, g_{\beta_n\nu} \partial_\mu \Psi + g_3\,\gamma_5\,\gamma_\mu\,\gamma^\rho\,\epsilon_{\rho\nu\alpha\beta_n}\partial^\alpha \Psi 
 \nonumber\\
& & + g_4\,\gamma_\mu\,\gamma^\rho\,(\partial_\rho g_{\nu\beta_n}  
- \partial_\nu g_{\rho\beta_n}) \Psi \bigr\} \,\partial_{\beta_1} \cdots\partial_{\beta_{n-1}} F^{\mu\nu} 
 \nonumber\\
& & +  \mathrm{H.c.}~,
\label{eq:L_electro}
\end{eqnarray}
respectively, where $\Psi$ is the field of the $\Sigma$ particle, $F^{\mu\nu}$ is the antisymmetric 
tensor of photon  field, 
and $\bar{\Psi}^{\mu_1\cdots\mu_n}$ is the modified RS-field of spin-$(n+1/2)$ particles 
constructed to make the interaction consistent as proposed in Ref.~\cite{Vrancx:2011qv}. 
The modified RS-field reads 
\begin{equation}
\bar{\Psi}_{\mu_1\cdots\mu_n} = O^{n+1/2}_{(\mu_1 \cdots \mu_n , \nu_1 \cdots \nu_n)\lambda_1 \cdots \lambda_n} (\partial) \bar{\psi}^{\lambda_1\cdots\lambda_n} \gamma^{\nu_1} \cdots \gamma^{\nu_n}~,
\end{equation}
where $\bar{\psi}$ is the original RS-field for spin-$(n+1/2)$ particles and the interaction operator $O$ 
is defined by
\begin{eqnarray}
&&O^{n+1/2}_{(\mu_1 \cdots \mu_n \nu_1 \cdots \nu_n)\lambda_1 \cdots \lambda_n} (\partial) ~=  \nonumber\\
&& \frac{1}{(n!)^2}\sum_{P(\nu)}\sum_{P(\lambda)} O^{3/2}_{(\mu_1 ,\nu_1)\lambda_1} \cdots O^{3/2}_{(\mu_n ,\nu_n)\lambda_n} ,
\end{eqnarray}
where $P(\nu)$ and $P(\lambda)$ indicate the permutations of all possible $\nu$ and $\lambda$ indices, respectively, and
\begin{equation}
O^{3/2}_{(\mu,\nu)\lambda}=(\partial_\mu g_{\nu\lambda} - \partial_\nu g_{\mu\lambda})~.
\end{equation}
The propagator used for calculating the scattering amplitude is obtained from the completeness relation of the RS-fields. 
This propagator, however, bears unphysical lower spin projection operator and eventually would yield unphysical contribution 
to the scattering amplitude if it was not properly handled. In this work, by using the interaction Lagrangians given by 
Eqs.~(\ref{eq:L_hadronic}) and (\ref{eq:L_electro}), the remaining lower spin terms are automatically removed from the
amplitude, leaving only the pure spin-$(n+1/2)$ contribution that comes from the projection operator 
\begin{equation}
\mathcal{P}^{n+1/2}_{\mu_1\cdots\mu_n ; \nu_1\cdots\nu_n} (p) = \frac{n+1}{2n+3} \gamma^\mu \mathcal{P}^{n+1}_{\mu\mu_1\cdots\mu_n ; \nu\nu_1\cdots\nu_n}(p)\gamma^\nu~,
\end{equation}
with the projection operator for spin-$n$ particle
\begin{eqnarray}
&&\mathcal{P}^{n}_{\mu_1\cdots\mu_n; \nu_1\cdots\nu_n}(p) \, =\,\frac{1}{n!^2}\sum_{P(\mu)}\sum_{P(\nu)}\sum_{k=0}^{k_\mathrm{max}} A^n_k \, 
\mathcal{P}_{\mu_1\mu_2} \nonumber\\
& &\times \mathcal{P}_{\nu_1\nu_2} \cdots \mathcal{P}_{\mu_{2k-1}\mu_{2k}}\mathcal{P}_{\nu_{2k-1}\nu_{2k}}\prod^n_{i=2k+1}\mathcal{P}_{\mu_i\nu_i}~,
\end{eqnarray}
where $k_\mathrm{max}$ is equal to $n/2$ if $n$ is even and to $(n-1)/2$ if $n$ is odd, 
$\mathcal{P}_{\mu\nu}(p)=(-g_{\mu\nu} + p_\mu \,p_\nu/p^2)$, and the coefficient $A^n_k$ is defined by 
\begin{equation}
A^n_k=\frac{(-1)^n}{(-2)^k}\frac{n!}{k!(n-2k)!}\frac{(2n-2k-1)!!}{(2n-1)!!}~.
\end{equation}

\begin{table*}[!]
\caption{Sources, types, channels, and number of experimental data used in the present analysis.}
\label{tab:data} 
\begin{ruledtabular}
\begin{tabular}{l l c r c c}
      Collaboration & Observable & Symbol & $N$ & Channel & Reference  \\
      \hline
      LEPS 2003 & Photon asymmetry & $\Sigma$ & 30 & $\gamma p\to K^+\Sigma^0$ & \cite{Zegers:2003ux} \\
      CLAS 2004 & Differential cross section & $d\sigma/d\Omega$ & 676 & $\gamma p\to K^+\Sigma^0$ & \cite{McNabb:2003nf} \\
                & Recoil polarization & $P$ & 146 & $\gamma p\to K^+\Sigma^0$ & \cite{McNabb:2003nf} \\
    SAPHIR 2004 & Differential cross section & $d\sigma/d\Omega$ & 480 & $\gamma p\to K^+\Sigma^0$ & \cite{Glander:2003jw} \\
                & Recoil polarization & $P$ & 12 & $\gamma p\to K^+\Sigma^0$ & \cite{Glander:2003jw} \\
      CLAS 2006 & Differential cross section & $d\sigma/d\Omega$ & 1280 & $\gamma p\to K^+\Sigma^0$ & \cite{Bradford:2005pt} \\
      LEPS 2006 & Differential cross section & $d\sigma/d\Omega$ & 39, 52 & $\gamma p\to K^+\Sigma^0$ & \cite{Sumihama:2005er,Kohri:2006yx} \\
                & Photon asymmetry & $\Sigma$ & 25, 26 & $\gamma p\to K^+\Sigma^0$ & \cite{Sumihama:2005er,Kohri:2006yx} \\
                & Differential cross section & $d\sigma/d\Omega$ & 72 & $\gamma n\to K^+\Sigma^-$ & \cite{Kohri:2006yx} \\
                & Photon asymmetry & $\Sigma$ & 36 & $\gamma n\to K^+\Sigma^-$ & \cite{Kohri:2006yx} \\
    SAPHIR 2006 & Differential cross section & $d\sigma/d\Omega$ & 90 & $\gamma p\to K^0\Sigma^+$ & \cite{Lawall:2005np} \\
                & Recoil polarization & $P$ & 10 & $\gamma p\to K^0\Sigma^+$ & \cite{Lawall:2005np} \\
      CLAS 2007 & Beam-Recoil polarization & $C_x$ & 94 & $\gamma p\to K^+\Sigma^0$ & \cite{Bradford:2006ba} \\
                & Beam-Recoil polarization & $C_z$ & 94 & $\gamma p\to K^+\Sigma^0$ & \cite{Bradford:2006ba} \\
     GRAAL 2007 & Recoil polarization & $P$ & 8 & $\gamma p\to K^+\Sigma^0$ & \cite{Lleres:2007tx} \\
                & Photon asymmetry & $\Sigma$ & 42 & $\gamma p\to K^+\Sigma^0$ & \cite{Lleres:2007tx} \\
      CLAS 2010 & Differential cross section & $d\sigma/d\Omega$ & 2089 & $\gamma p\to K^+\Sigma^0$ & \cite{Dey:2010hh} \\
                & Recoil polarization & $P$ & 455 & $\gamma p\to K^+\Sigma^0$ & \cite{Dey:2010hh} \\
                & Differential cross section & $d\sigma/d\Omega$ & 177 & $\gamma n\to K^+\Sigma^-$ & \cite{AnefalosPereira:2009zw} \\
  Crystal Ball 2014 & Differential cross section & $d\sigma/d\Omega$ & 1129 & $\gamma p\to K^+\Sigma^0$ & \cite{Jude:2013jzs} \\
      CLAS 2016 & Recoil polarization & $P$ & 127 & $\gamma p\to K^+\Sigma^0$ & \cite{Paterson:2016vmc} \\
                & Photon asymmetry & $\Sigma$ & 127 & $\gamma p\to K^+\Sigma^0$ & \cite{Paterson:2016vmc} \\
                & Target asymmetry & $T$ & 127 & $\gamma p\to K^+\Sigma^0$ & \cite{Paterson:2016vmc} \\
                & Beam-Recoil polarization & $O_x$ & 127 & $\gamma p\to K^+\Sigma^0$ & \cite{Paterson:2016vmc} \\
                & Beam-Recoil polarization & $O_z$ & 127 & $\gamma p\to K^+\Sigma^0$ & \cite{Paterson:2016vmc} \\
        MAMI A2 2018 & Differential cross section & $d\sigma/d\Omega$ & 39 & $\gamma p\to K^0\Sigma^+$ & \cite{Akondi:2018shh} \\
                & Differential cross section & $d\sigma/d\Omega$ & 48 & $\gamma n\to K^0\Sigma^0$ & \cite{Akondi:2018shh} \\
      \hline
      \multicolumn{3}{l}{Total number of data} & 7784 &  & \\
\end{tabular}
\end{ruledtabular}
\end{table*}

As mentioned above the constructed scattering amplitude is free from the unphysical lower-spin contribution 
that originates from the RS-fields. In the compact form the amplitude can be written as
\begin{eqnarray}
\mathcal{M}^{R}_{\mathrm{fi}}&=& \bar{u}_\Lambda\,\Gamma_{\mu_1\cdots\mu_n}^\mathrm{had}\, p_R^{2n}\,\frac{\slashed{p}_R + m_{N^*}}{p_R^2-m_{N^*}^2+im_{N^*}\Gamma} \nonumber\\
& &\times \mathcal{P}^{\mu_1\cdots\mu_n\,,\,\nu_1\cdots\nu_n}_{(n+1/2)}(p_R)\,\Gamma_{\nu_1\cdots\nu_n}^\mathrm{em}u_p~.
\label{eq:amplitude_Mfi}
\end{eqnarray}
where $p_R$ is the four-momentum of resonance particle and the vertex factors $\Gamma_{\mu_1\cdots\mu_n}^\mathrm{had}$
and $\Gamma_{\nu_1\cdots\nu_n}^\mathrm{em}$ are derived directly from the interaction Lagrangians given by 
Eqs.~(\ref{eq:L_hadronic}) and (\ref{eq:L_electro}), respectively. For the purpose of numerical computation
of observables we need to calculate the total scattering amplitude, which is obtained by adding the 
background $\mathcal{M}^{\rm back}_{\mathrm{fi}}$ and resonance 
$\mathcal{M}^{R}_{\mathrm{fi}}$ contributions, i.e.,
\begin{eqnarray}
\mathcal{M}_{\mathrm{fi}} &=& \mathcal{M}^{\rm back}_{\mathrm{fi}} + \sum_{R} \mathcal{M}^{R}_{\mathrm{fi}} ~,
\label{eq:amplitude_total}
\end{eqnarray}
where the summation on the right hand side is performed over all nucleon resonances considered in the
present work. The total amplitude 
is decomposed into the gauge and Lorentz invariant matrices $M_{i}$ 
\begin{eqnarray}
{\cal M}_{\mathrm{fi}} &=& {\bar u}_\Sigma \sum_{i=1}^4 A_{i}(s,t,u)\, M_{i}\, u_N~ , 
\label{eq:scattering-amplitudes-Mi}
\end{eqnarray}
where $M_i$ denotes the gauge and Lorentz invariant matrices given by \cite{Clymton:2017nvp,Mart:2019fau}
\begin{eqnarray}
\label{eq:M1}
M_{1} & = & \gamma_{5}\, \epsilon\!\!/ k\!\!\!/ \, ,\\
\label{eq:M2}
M_{2} & = & \gamma_{5}\left( 2q \cdot \epsilon\, P \cdot k - 2q 
\cdot k\, P \cdot \epsilon \right)~ ,\\
\label{eq:M3}
M_{3} & = & \gamma_{5} \left( q\cdot k\, \epsilon\!\!/ - q\cdot \epsilon \,
k\!\!\!/ \right) ~ ,\\
\label{eq:M4}
M_{4} & = & i \epsilon_{\mu \nu \rho \sigma} \gamma^{\mu} q^{\nu}
\epsilon^{\rho} k^{\sigma}~ ,
\end{eqnarray}
with $P = \frac{1}{2}(p + p_{\Sigma})$ and 
$\epsilon_{\mu \nu \rho \sigma}$ is the four dimensional 
Levi-Civita tensor. The required cross section and polarization observables
can be calculated from the functions $A_i$ given by Eq.~(\ref{eq:scattering-amplitudes-Mi}), 
which depend on the Mandelstam variables given in Eq.~(\ref{eq:Mandelstam}).

Note that as seen from Eq.~(\ref{eq:amplitude_Mfi}) the present formalism introduces the 
high momentum dependence $p_R^{2n}$ that might lead to a non-resonance behavior of scattering 
amplitude at high energies. To alleviate this problem we need a stronger hadronic form factor 
that can sufficiently suppresses the divergence of the amplitude at high energies. The widely use 
form factor is the dipole one, i.e., 
\begin{equation}
F_\mathrm{had} = \frac{\Lambda^4}{(s-M^2)^2+\Lambda^4}~,
\label{eq:ff_dipole_n1}
\end{equation}
with $\Lambda$ the cutoff parameter and $M$ the resonance mass. However, we found that such a form factor does 
not have a sufficiently strong suppression for this purpose. Therefore, in this study we propose the
use of the multi-dipole form factor 
\begin{equation}
F_\mathrm{had} = \left\{\frac{\Lambda^4}{(s-M^2)^2+\Lambda^4}\right\}^n~,
\label{eq:ff_dipole_n}
\end{equation}
as well as the Gaussian one 
\begin{equation}
F_\mathrm{had} = \exp\left\{-(s-M^2)^2/\Lambda^4\right\} ~,
\label{eq:ff_dipole_gauss}
\end{equation}
and investigate the effects of these form factors on the constructed model. 
For the multi-dipole form factor we will present the result with $n=3$, for 
which we obtained the best agreement with experimental data, and denote the 
model with  HFF-P3. The models that use the form factors given by 
Eqs.~(\ref{eq:ff_dipole_n1}) and (\ref{eq:ff_dipole_gauss}) are denoted with  
HFF-P1 and  HFF-G, respectively.

Equations~(\ref{eq:L_hadronic}) and (\ref{eq:L_electro}) indicate that for each resonance with $J\geq 3/2$ there 
are four unknown coupling constants which might be considered as free parameters. These coupling constants
can be extracted from fitting the calculated observables to the corresponding experimental data. In the present 
work the fitting process was performed by using the {\small CERN-MINUIT} code \cite{James:1975dr} 
to minimize the value of 
\begin{eqnarray}
  \label{chi2def}
  \frac{\chi^2}{N_{\rm dof}} &=& \frac{1}{N_{\rm data}-N_{\rm par}} 
  \sum_{i=1}^{N_{\rm data}}~\left[\frac{\sigma_i^{\rm exp} - 
  \sigma_i^{\rm th}}{\Delta \sigma_i^{\rm exp}} \right]^2 ~,
\end{eqnarray}
where $N_{\rm data}$ and $N_{\rm par}$ indicate the numbers of experimental data and free
parameters  used in the fit, respectively, $\sigma_i^{\rm exp}$ and $\sigma_i^{\rm th}$ are the $i$-th
values of experimental and theoretical observables, and $\Delta \sigma_i^{\rm exp}$ is the  corresponding
experimental error bar. 

Experimental data used in the present analysis are obtained from a number of 
experimental collaborations as listed in Table~\ref{tab:data}. Note that the $K\Sigma$ 
photoproduction offers 4 possible isospin channels.  Among them the 
$\gamma p \to K^+\Sigma^0$ channel 
has the largest number of experimental data, as can be seen  in Table~\ref{tab:data}.
Furthermore, experimental data of this channel are available for different types of
observables, which can be expected to complete our understanding of this reaction. 
Nevertheless, although with a limited number and observable types, the existence of 
experimental data in the other three isospin channels is very important to constrain 
the extracted coupling constants as well as the predicted observables \cite{Mart:1995wu}.

\subsection{Resonances Properties}
In this study, we extract a number of important resonance  properties, i.e., their masses and 
total widths evaluated at pole, their partial widths, and their individual contribution to the process. 
The evaluation of mass and width at pole starts with a complex root equation of the denominator of 
the scattering amplitude, which reads
\begin{equation}
\label{eq:root_sp}
s_\mathrm{p} - m_R^2 + i m_R \Gamma_R = 0 ~,
\end{equation}
where $m_R$ and $\Gamma_R$ are the Breit-Wigner mass and width, respectively. The variable $s_{\rm p}$ is defined as
\begin{equation}
s_\mathrm{p} = (m_\mathrm{p} - i \Gamma_\mathrm{p}/2)^2
\end{equation}
where $m_\mathrm{p}$ and $\Gamma_\mathrm{p}$ are the mass and width evaluated at the pole, respectively. 
We can clearly see that the solutions of $m_\mathrm{p}$ and $\Gamma_\mathrm{p}$ were actually simple. 
However, in the present work, we use an energy-dependent width $\Gamma (s)$ that is directly proportional 
to the total width $\Gamma_R$~\cite{Lee:2001}. As a consequence, we cannot analytically solve 
the root equation given in Eq.~(\ref{eq:root_sp}). Therefore, in the present analysis we solve this
equation numerically.

Furthermore, we can also compute the partial decay widths $\sqrt{\Gamma_{\gamma p}\Gamma_{K\Sigma}}/\Gamma_{\mathrm{tot}}$
to conveniently compare the values of extracted coupling constants in this study. To this end, we start 
with the interaction Lagrangians for spin $J=n+1/2$ resonances and for each $J$ we obtain 
the decay width formula. Note that in the present analysis we can only obtain the product of hadronic 
and electromagnetic decay widths, because in the single channel analysis only the product of the electromagnetic 
and hadronic coupling constants can be extracted. The formulation of these decay widths can be found in 
our previous study~\cite{Clymton:2017nvp}.

The significance of each resonance can be also evaluated by excluding the specific resonance 
in the fitting process. Mathematically, the significance of an $N^*$ resonance is defined through 
\begin{eqnarray}
  \label{eq:par_res}
  \Delta \chi^2 &=& \frac{\chi^2_{{\rm all}-N^*}-\chi^2_{\rm all}}{\chi^2_{\rm all}}
  \times 100\,\% ~,
\end{eqnarray}
where $\chi^2_{\rm all}$ is obtained from fitting the data by using 
all resonances, while $\chi^2_{{\rm all}-N^*}$ is obtained by using all 
but a specific $N^*$ resonance. In the present work Eq.~(\ref{eq:par_res}) 
is also used for investigating the significance of $\Delta$ resonances. 
The significance of a resonance is not only very useful for investigation of  
the role of each resonance in the reaction, but also for a practical 
guidance to simplify the model if minimizing the number of resonances used in the 
model is important. A simple covariant isobar model is very important for 
the application in few-body nuclear physics, for which numerical computation 
and accuracy are very demanding. 

\section{RESULTS AND DISCUSSION}
\label{sec:result}

\subsection{General Results}
\begin{figure}[!]
\includegraphics[scale=0.47]{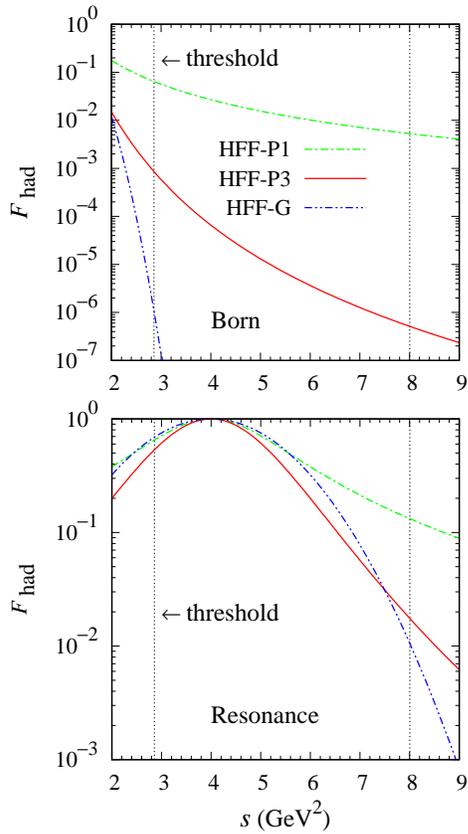}
\caption{The Born and resonance hadronic form factors given by Eqs.~(\ref{eq:ff_dipole_n1})-(\ref{eq:ff_dipole_gauss})
  with the cutoffs extracted from fitting to experimental data. In the bottom panel, 
  for the sake of visibility, the mass of resonance is chosen to be 2000 MeV.
  The energies covered by the experimental data used in the present work are limited by the two vertical 
  dotted lines shown in each panel.}
\label{fig:hff}
\end{figure}

Table~\ref{tab:chi2n} compares the extracted coupling constants of the Born terms obtained
for the 3 different hadronic form factor models. By comparing the values of $\chi^2$, we 
can clearly see that the model HFF-P1 yields less  agreement with experimental data. 
This indicates that to achieve a better agreement with experimental data we have to
use a softer form factor. The result obtained from model HFF-P3 corroborates this. However,
since the three form factors used in the present analysis have different forms, see 
Eqs.~(\ref{eq:ff_dipole_n1})-(\ref{eq:ff_dipole_gauss}), it is difficult to estimate 
the suppression imposed by the form factors on the amplitude by merely comparing their 
cutoffs listed in Table~\ref{tab:chi2n}. Therefore, in Fig.~\ref{fig:hff} we plot the 
form factors for both Born and resonance terms as a function of the Mandelstam variable $s$
and shows the energy region covered by the experimental data in the fitting database.
From the top panel of Fig.~\ref{fig:hff} we may conclude that the Gaussian form factor is the softest
one and, as a consequence, this form factor strongly suppresses the Born amplitude. In contrast to this 
form factor, the dipole one given by Eq.~(\ref{eq:ff_dipole_n1}) provides the lightest 
suppression, whereas the multi-dipole form factor yields the relatively moderate suppression.
Nevertheless, even the dipole form factor decreases the contribution of Born terms 
significantly. Thus, from Fig.~\ref{fig:hff} we may safely say that all models analyzed 
in this work are resonance-dominated models because, as shown in the bottom panel  of 
Fig.~\ref{fig:hff}, near the resonance mass the form factors do not suppress the amplitude
significantly. 

\begin{table}[!]
\caption{Background parameters and the $\chi^2$ contributions from individual isospin channels   
  obtained from all models investigated in the present work.}
\label{tab:chi2n} 
\begin{ruledtabular}
  \begin{tabular}{lrrr}
      Parameters & HFF-P1 & HFF-P3 & HFF-G  \\
      \hline
      $g_{K \Lambda N} / \sqrt{4 \pi}$ & $-3.00$  & $-3.00$ & $-3.00$ \\
      $g_{K \Sigma N} / \sqrt{4 \pi}$  & 0.90     & 1.30 & 1.30 \\
      $G^{V}_{K^*} / 4\pi$             &  $-0.15$ & $-0.05$ & $-0.13$ \\
      $G^{T}_{K^*} / 4\pi$             &  $-0.21$ &   0.11  &   0.22  \\
      $G^{V}_{K_1} / 4\pi$             &    0.12  & $-0.28$ & $-0.26$ \\
      $G^{T}_{K_1} / 4\pi$             &    4.37  &   0.45  & $-0.54$ \\
      $\Lambda_{\rm B}$(GeV)           &  0.72    &   0.80  &   0.73  \\
      $\Lambda_{\rm R}$(GeV)           &  1.25    &   1.54  &   1.37  \\
      $\theta_{\rm had}$(deg)          & 90.0     &   90.0  &   53.4  \\
      $\phi_{\rm had}$(deg)            &   0.00   &   0.00  &  180.0   \\
      \hline 
      $\chi^2_{K^+\Sigma^0}$ &  8657  &  8259  &  8282 \\
      $\chi^2_{K^0\Sigma^+}$ &   221  &   190  &   198 \\
      $\chi^2_{K^+\Sigma^-}$ &   158  &   146  &   153 \\
      $\chi^2_{K^0\Sigma^0}$ &    17  &    20  &    17 \\
      $\chi^2 / N$           &  1.22  &  1.16  &  1.16 \\
\end{tabular}
\end{ruledtabular}
\end{table}

\begin{figure}[!]
\includegraphics[scale=0.5]{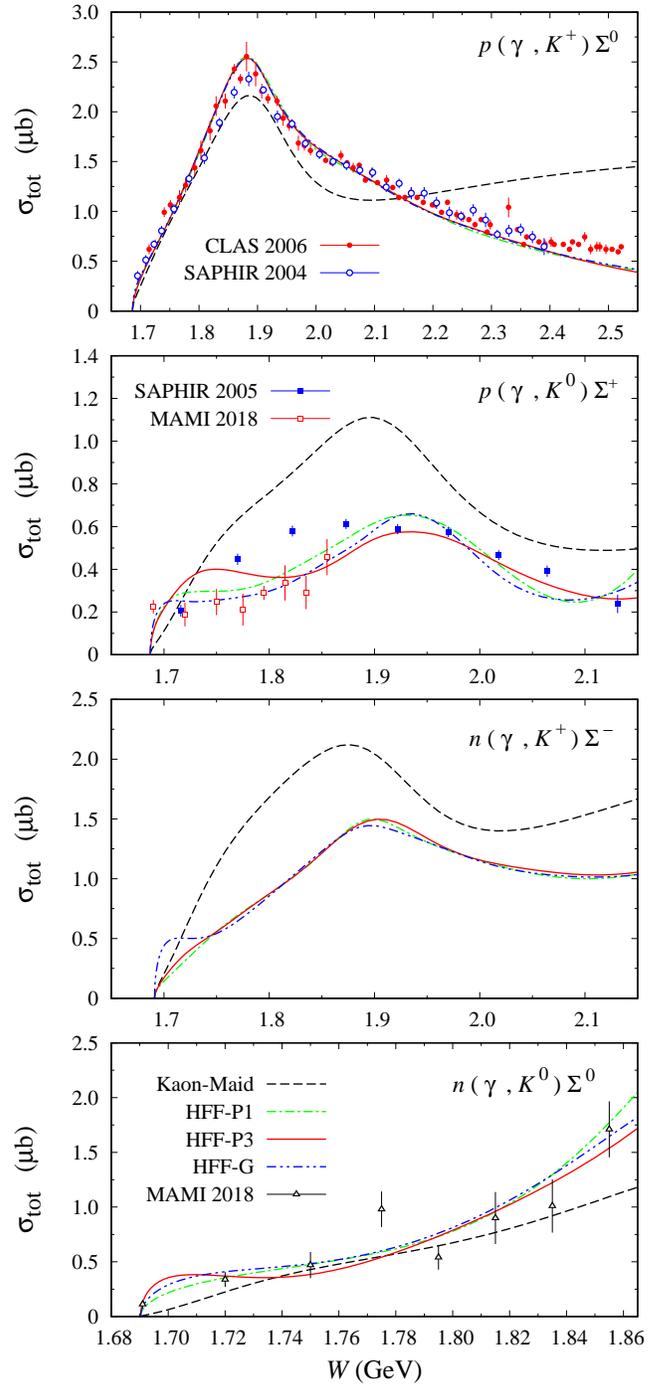}
\caption{Calculated total cross sections of the $\gamma p\to K^+\Sigma^0$, $\gamma p\to K^0\Sigma^+$, $\gamma n\to K^+\Sigma^-$, and 
$\gamma n\to K^0\Sigma^0$ isospin channels obtained from Kaon-Maid \cite{missing-d13} and different models analyzed in the present 
work (HFF-P1, HFF-P3, and HFF-G), compared with the presently available experimental data. Notation of the curves is given in 
the lowest panel, whereas notation of the data is given in the corresponding panels. Note that the experimental data shown 
in this figure were not included in the fitting process of the present work.}
\label{fig:kstot}
\end{figure}

\begin{figure*}[!]
\includegraphics[scale=0.8]{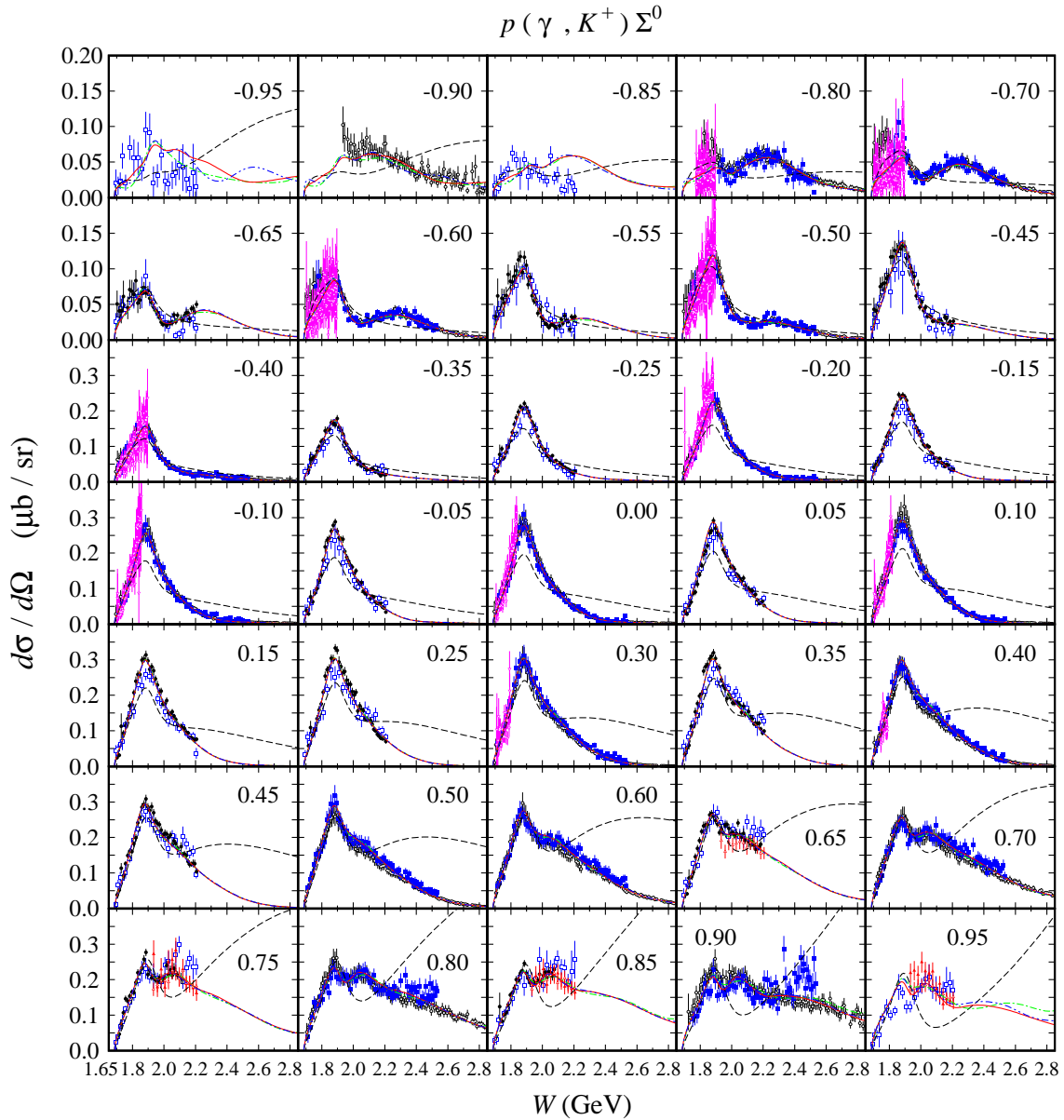}
\caption{Comparison between experimental data and calculated differential cross sections as a function of the total c.m. energy
  $W$ for the $\gamma p \to K^+\Sigma^0$ isospin 
  channel. Notation of the curves is as in Fig.~\ref{fig:kstot}. The corresponding value of $\cos\theta$ is denoted in each
  panel. Experimental data are obtained 
  from the CLAS 2004 (solid circles~\cite{McNabb:2003nf}), SAPHIR 2004 (open squares~\cite{Glander:2003jw}), CLAS 2006 
  (solid square~\cite{Bradford:2005pt}), LEPS 2006 (solid~\cite{Sumihama:2005er} and open~\cite{Kohri:2006yx} triangles), 
  CLAS 2010 (open circles~\cite{Dey:2010hh}), and Crystal Ball 2014 (open inverted-triangles~\cite{Jude:2013jzs}) collaborations.}
\label{fig:difcs3w}
\end{figure*}

\begin{figure*}[!]
\includegraphics[scale=0.70]{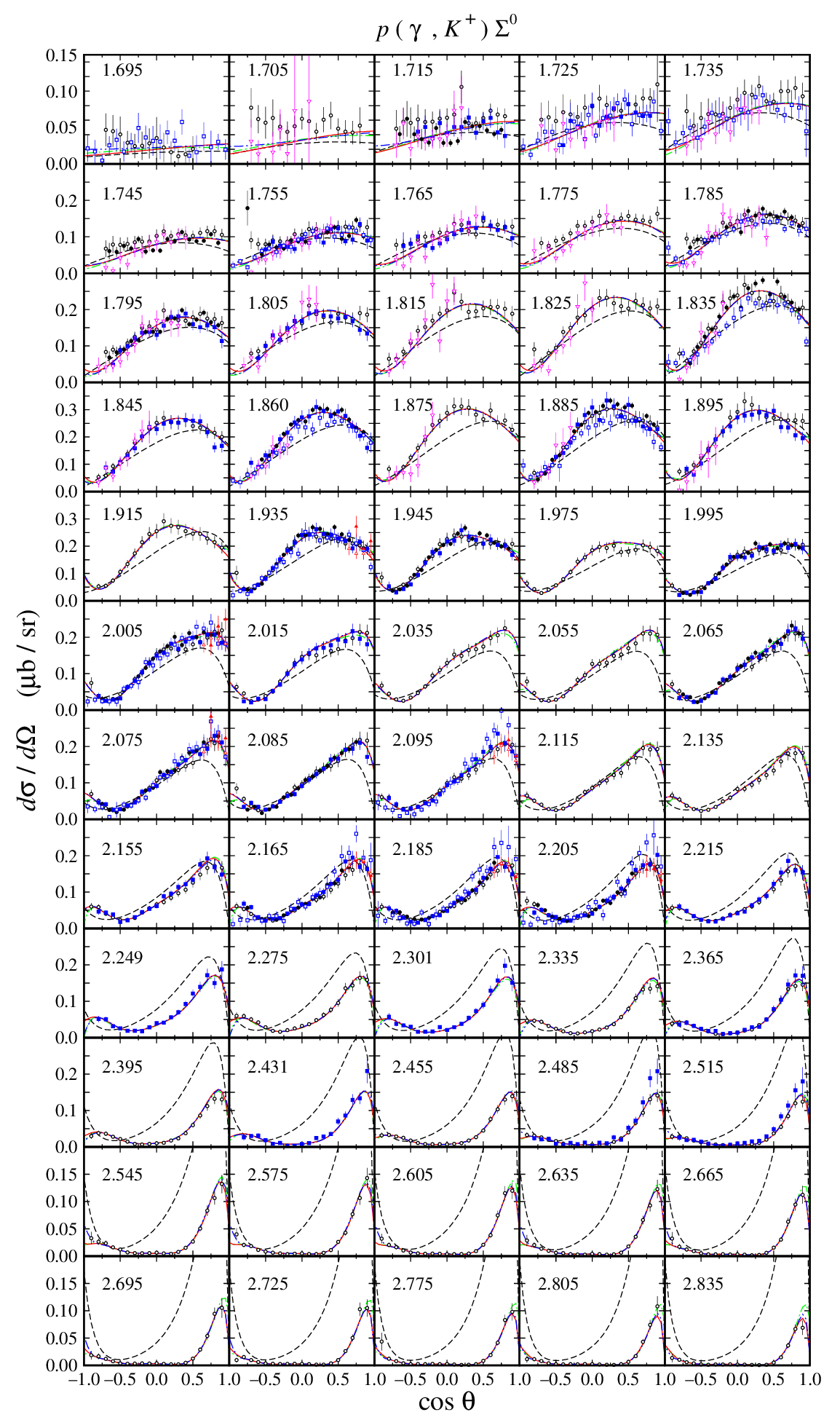}
\caption{As in Fig.~\ref{fig:difcs3w}, but for the angular distribution of differential cross section. 
  The corresponding value of the total c.m. energy $W$ in GeV is given in each panel.}
\label{fig:difcs3th}
\end{figure*}

In the case of resonance, within the covered energy the suppression effect 
of form factor is clearly not symmetrical.
This is understandable because the resonance masses used in the present analysis 
are below 2.3 GeV. However, this asymmetrical
suppression is required for the covariant description of a resonance due to the large
contribution of a Z-diagram that indicates the existence of a particle and an antiparticle 
in the intermediate state \cite{Mart:2019jtb}. This contribution increases quickly as
the energy of resonance increases beyond the resonance mass and, therefore, requires 
an increasing suppression. In our previous study  \cite{Mart:2019jtb} it was shown 
that the dipole form factor given by Eq.~(\ref{eq:ff_dipole_n1}) is suitable for
this purpose.

In Table~\ref{tab:chi2n} we also show the $\chi^2$ contribution from each channel. 
It is apparent from this table that the model HFF-P3 shows the best
agreement with the experimental data (lowest $\chi^2$) from all but the $\gamma n \to K^0\Sigma^0$ 
channel. As we will see later, when we compare the observables, the model deficiency 
in this channel originates from the discrepancy between the calculated differential cross section 
and the experimental data at forward angles. Nevertheless, the effect of this discrepancy is less 
significant compared to those obtained from the other three isospin channels.
Furthermore, by analyzing the sources of experimental data we found that the model HFF-P3 
yields a nice agreement not only with the SAPHIR \cite{Lawall:2005np}, but also with the 
MAMI \cite{Akondi:2018shh} data. 

The calculated total cross sections obtained from the three models are compared 
with the available experimental data in Fig.~\ref{fig:kstot}. Note that 
although there are no data for the $K^+\Sigma^-$ total cross section, 
both LEPS 2006 \cite{Kohri:2006yx} and CLAS 2010 \cite{AnefalosPereira:2009zw}
collaborations have measured the $K^+\Sigma^-$ 
differential cross section, whereas the LEPS 2006 collaboration has
obtained the $K^+\Sigma^-$ photon asymmetry \cite{Kohri:2006yx}. Since
these data have been included in our fitting database, we believe that
the calculated $K^+\Sigma^-$ total cross section shown in Fig.~\ref{fig:kstot} 
is also accurate.

From  Fig.~\ref{fig:kstot} we may conclude that all three models can nicely reproduce 
the total cross section data. It is understandable that this new result is quite 
different from that of Kaon-Maid, since Kaon-Maid was fitted to the old SAPHIR data 
\cite{saphir98,Goers:1999sw},
except for the $\gamma n \to K^0\Sigma^0$ reaction, for which Kaon-Maid yields a 
fairly good prediction to the recent MAMI data. We believe that the latter is 
pure coincidence and, in fact, the discrepancy between Kaon-Maid prediction and 
experimental data is clearly seen near the threshold 
and higher energy regions. It is also important to note that in the $K^0\Sigma^+$ 
channel the total cross section indicates two resonance peaks. Interestingly, 
they originate from the $N(1720)P_{13}$ and $N(1900)P_{13}$ states that have been
found to be important to describe both $K\Lambda$ and $K\Sigma$ photoproductions
\cite{Mart:2019fau,Mart:2000jv}. The first peak does not appear in the $K^0\Sigma^+$
channel, whereas the second one is shifted above 1.9 GeV due to the interference
with other resonances, whose extracted masses are heavier than 1.9 GeV.


\begin{figure*}[!]
\includegraphics[scale=0.8]{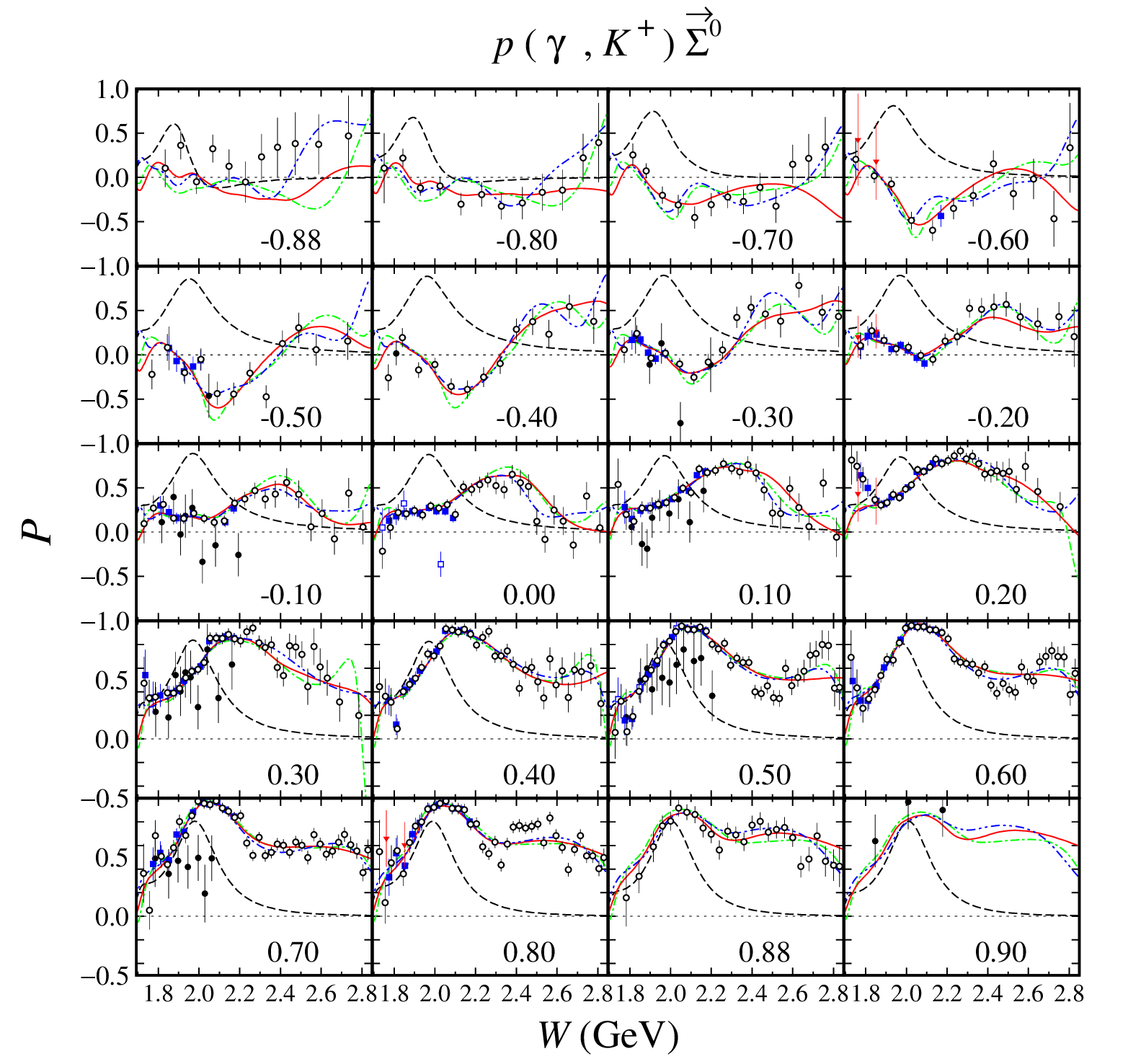}
\caption{Total c.m. energy distribution of the recoil polarization in the $\gamma p \to K^+ \vec{\Sigma}^0$ reaction 
obtained from all models shown in Fig.~\ref{fig:kstot}. Notation of the curves is as in Fig.~\ref{fig:kstot}. 
The corresponding value of $\cos\theta$ is given in each  panel.
Experimental data are obtained from the CLAS 2004 (solid circles~\cite{McNabb:2003nf}), SAPHIR 2004 (open 
squares~\cite{Glander:2003jw}), GRAAL 2007 (solid inverted-triangles~\cite{Lleres:2007tx}), CLAS 2010 
(open circles~\cite{Dey:2010hh}), and CLAS 2016 (solid squares~\cite{Paterson:2016vmc}) collaborations.}
\label{fig:recpo3w}
\end{figure*}
\begin{figure*}[!]
\includegraphics[scale=0.7]{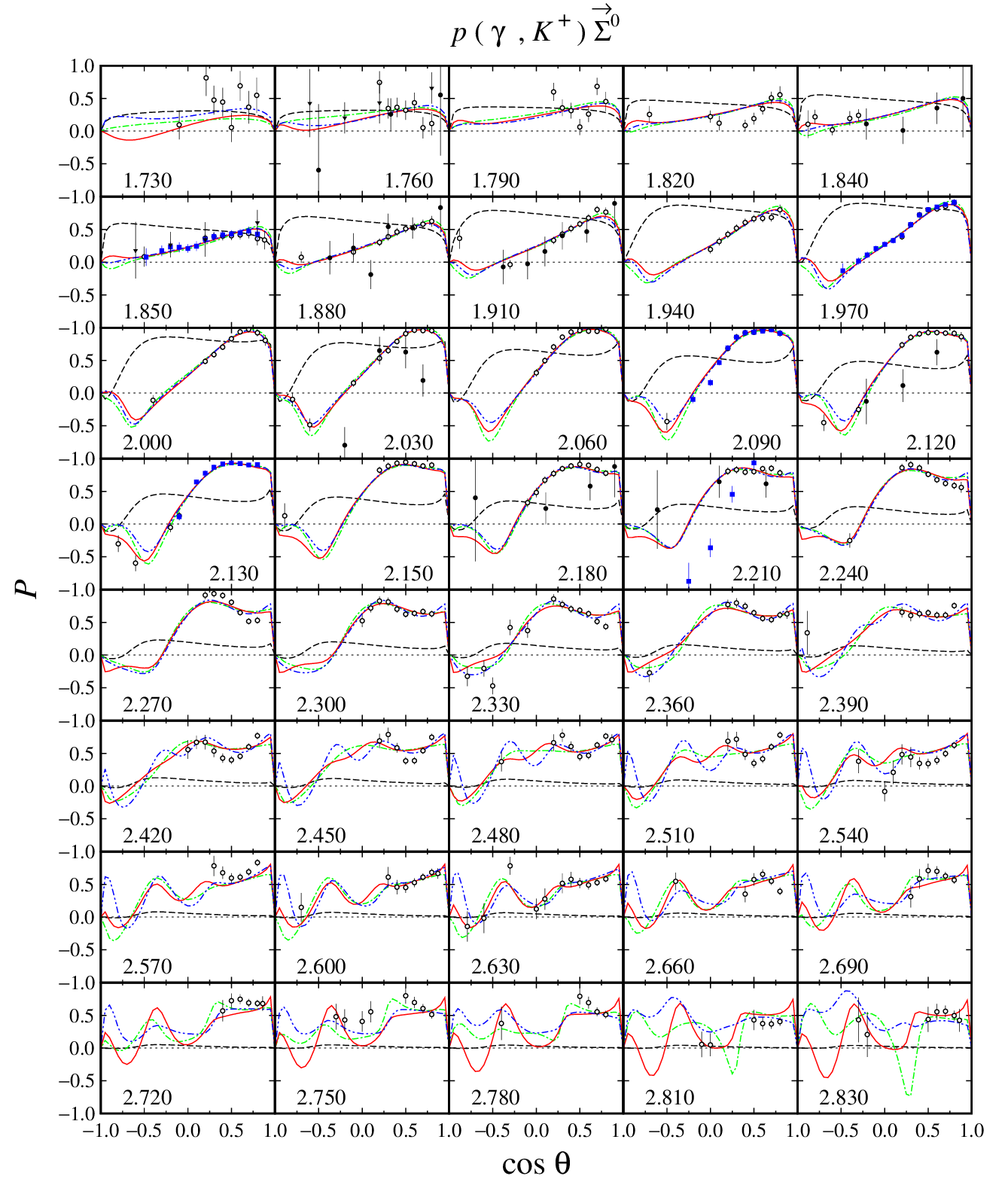}
\caption{As in Fig.~\ref{fig:recpo3w}, but for angular distribution.
       The corresponding value of the total c.m. energy $W$ in GeV is given in each panel.}
\label{fig:recpo3th}
\end{figure*}

From Figs.~\ref{fig:difcs3w} and \ref{fig:difcs3th} we can clearly see that all models do not 
exhibit significant variation in the differential cross sections, except for the extreme 
kinematics, i.e., in the forward and backward
directions and in the higher energy region. In this kinematics, the interference between the 
resonance and background terms is found stronger than in any other regions. Meanwhile, 
experimental data in this kinematic are scattered and, in fact, in certain energy 
intervals there are no data available to constrain the models. Therefore, during the
fitting process this condition yields significant variations among the models. 

The lack of experimental data in certain energy and angular regions also occurs 
in the case of polarization observables. It is well known that unlike the cross section, 
the polarization observables are very sensitive to the ingredient of the reaction 
amplitude. Therefore, they can severely constrain the flexibility of the model during the
fitting process. As a result, the calculated $\chi^2$ reported in the present work
originates mostly from the polarization data.

From their formulations it is easy to understand that the
single polarization observables are the simplest polarization observables 
that depend sensitively on all ingredients of the model, i.e., not only  
the resonance configuration, but also the background structure. Thus, 
they are very useful to constrain the models that predict similar 
trend in differential cross section, but significantly different 
beam, target, or recoil polarization observables. To this end,
we also note that there was a discussion on whether the right model 
should be resonance- or background-dominated, or both resonance 
and background are equally contributing 
\cite{missing-d13,Mart:2017xtf,Mart:ptep2019}. Hence, the observables
can be used to help alleviate the problem. 
Furthermore, in the certain kinematical region, where differential cross 
section data are not available, single polarization observables 
provide an important tool to shape the trend of differential cross section. 

\begin{figure*}[!]
\includegraphics[scale=0.8]{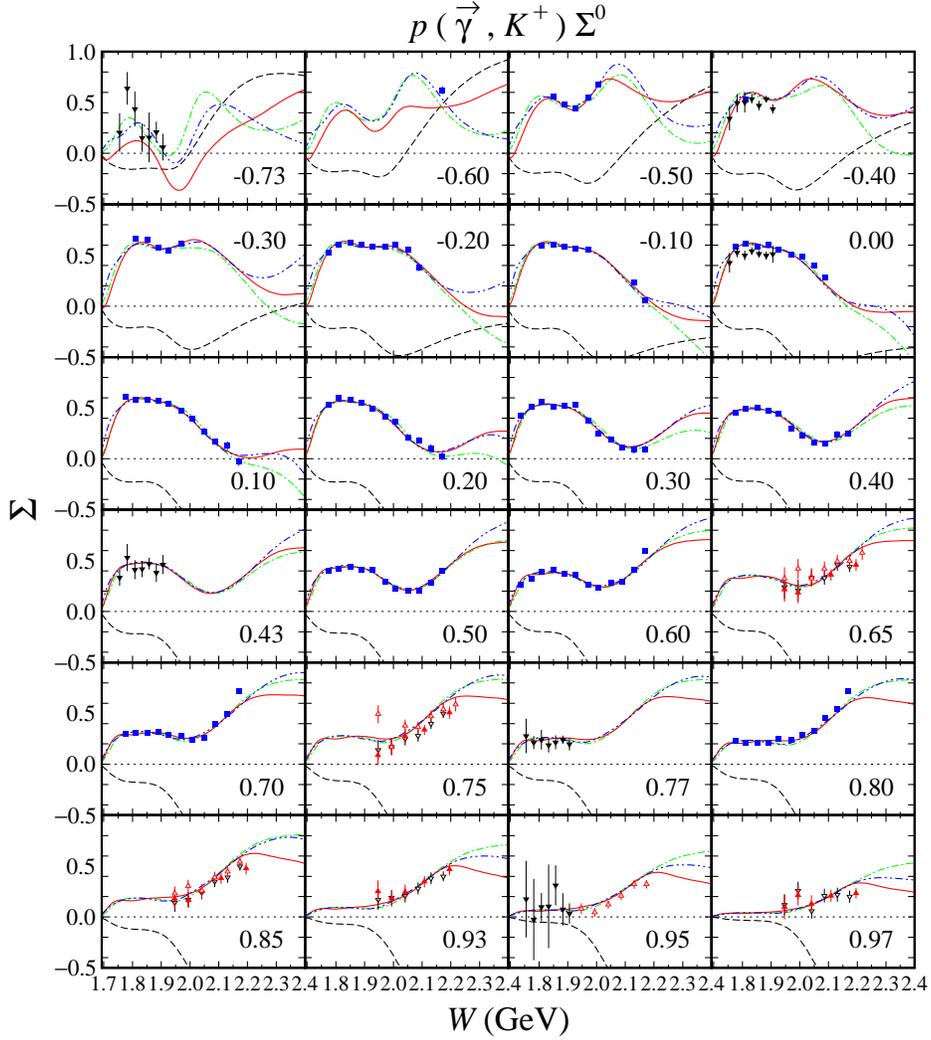}
\caption{Total c.m. energy distribution of the photon asymmetry in the 
$\vec{\gamma} p \to K^+ \Sigma^0$ reaction obtained from all models shown 
in Fig.~\ref{fig:kstot}. Notation of the curves is as in Fig.~\ref{fig:kstot}. The 
corresponding value of $\cos\theta$ is given in each  panel. Experimental data 
are obtained from the LEPS 2003 (open inverted-triangles~\cite{Zegers:2003ux}), 
LEPS 2006 (solid~\cite{Sumihama:2005er} and open~\cite{Kohri:2006yx} triangles), 
GRAAL 2007 (solid inverted-triangles~\cite{Lleres:2007tx}), and CLAS 2016 (solid 
squares~\cite{Paterson:2016vmc}) collaborations.}
\label{fig:phas3w}
\end{figure*}

\begin{figure*}[!]
\includegraphics[scale=0.8]{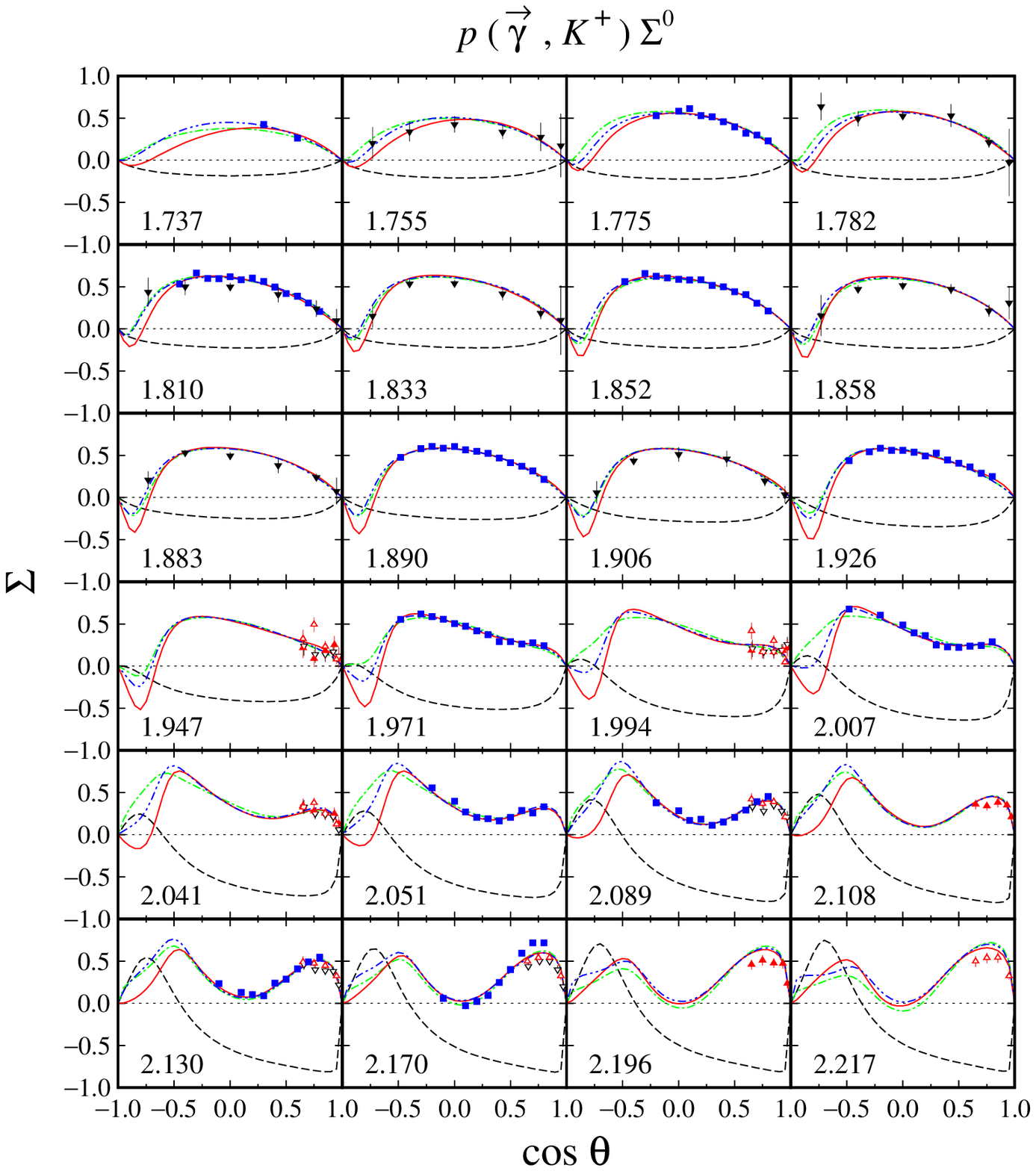}
\caption{As in Fig.~\ref{fig:phas3w}, but for angular distribution.
       The corresponding value of the total c.m. energy $W$ in GeV is given in each panel.}
\label{fig:phas3th}
\end{figure*}

\begin{figure}[!]
\includegraphics[scale=0.72]{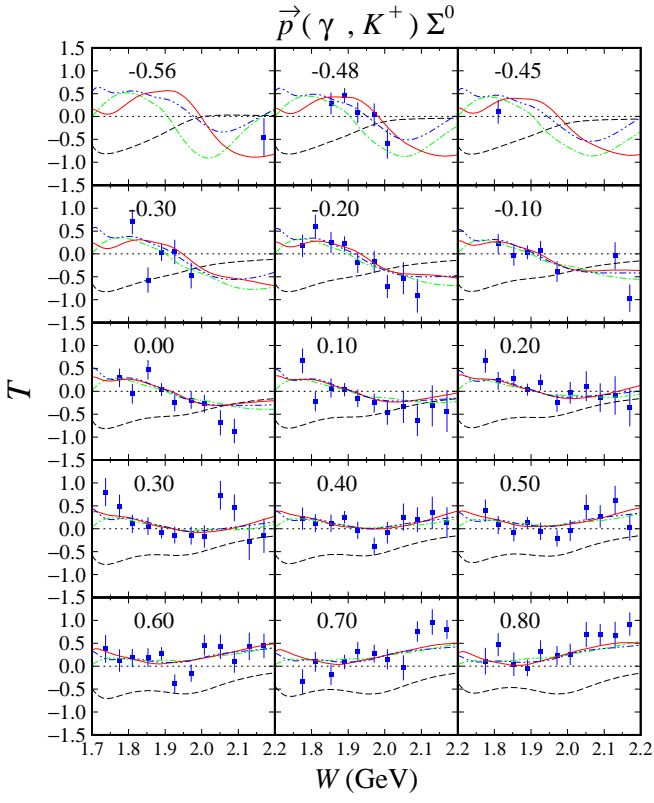}
\caption{Energy distribution of the $\gamma \vec{p} \to K^+ \Sigma^0$  target asymmetry 
  obtained from all models. Notation of the curves is as in Fig.~\ref{fig:kstot}. 
  Experimental data are obtained from the 
  CLAS 2016 collaboration~\cite{Paterson:2016vmc}.}
\label{fig:tarpo3w}
\end{figure}

\begin{figure}[!]
\includegraphics[scale=0.73]{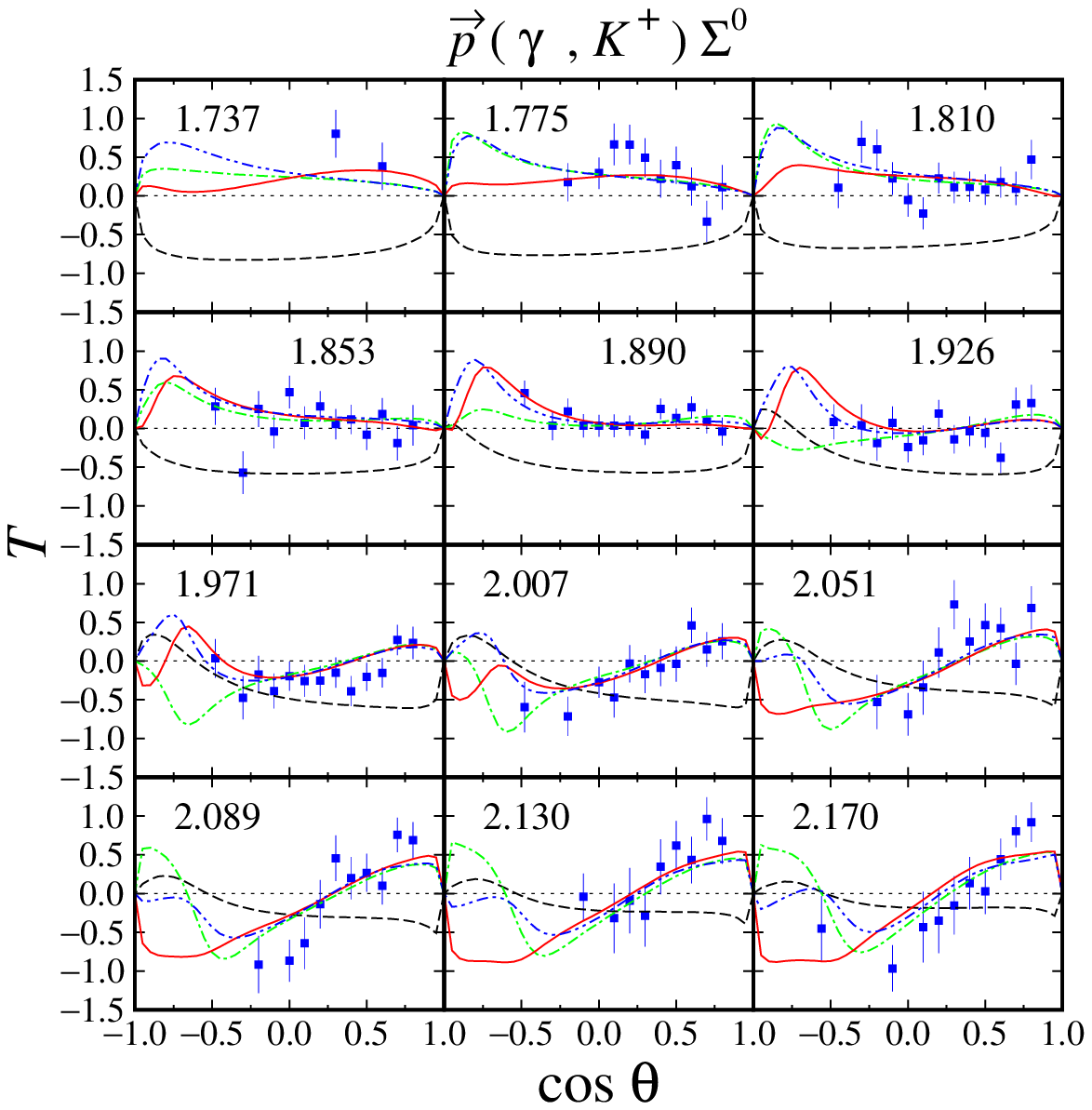}
\caption{As in Fig.~\ref{fig:tarpo3w}, but for angular distribution.}
\label{fig:tarpo3th}
\end{figure}

From Figs.~\ref{fig:recpo3w} and \ref{fig:recpo3th} we found that 
in the kinematical region where the differential cross section data 
are abundant, e.g., at $\cos\theta = -0.70$, the three models show a
large variation of the calculated recoil polarization $P$, in contrast to the 
predicted differential cross sections. On the other hand, in 
the kinematical region where differential cross section data 
are not available, e.g., at $\cos\theta = -0.95$, we find that 
the models yield significant variation in  both cross section 
and polarization observables. This phenomenon is also shown by 
the other single polarization observables, i.e., the photon asymmetry 
$\Sigma$ shown in Figs.~\ref{fig:phas3w} and \ref{fig:phas3th}, as well
as the target asymmetry $T$ shown in Figs.~\ref{fig:tarpo3w} and 
\ref{fig:tarpo3th}. Therefore, we may conclude that single polarization 
observables are potential to become important tools to determine the 
correct ingredients of the Born and resonance terms in the
$K\Sigma$ photoproduction channels.

Double-polarization observables are even more sensitive to the constituent 
of the reaction amplitude compared to the single polarization ones. Therefore, 
double-polarization observables can be considered as the main constraint to 
the models. 

\begin{figure}
\includegraphics[scale=0.8]{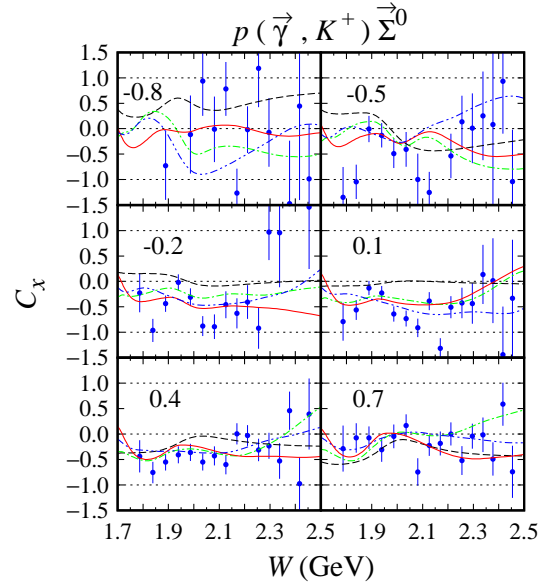}
\caption{Energy distribution of the beam-recoil polarization $C_x$ in 
the $\vec{\gamma} p \to K^+ \vec{\Sigma}^0$ reaction obtained from all models. 
Notation of the curves is  as in Fig.~\ref{fig:kstot}. Experimental data are 
obtained from the CLAS 2007 collaboration~\cite{Bradford:2006ba}.}

\label{fig:cx3w}
\end{figure}

\begin{figure}
\includegraphics[scale=0.73]{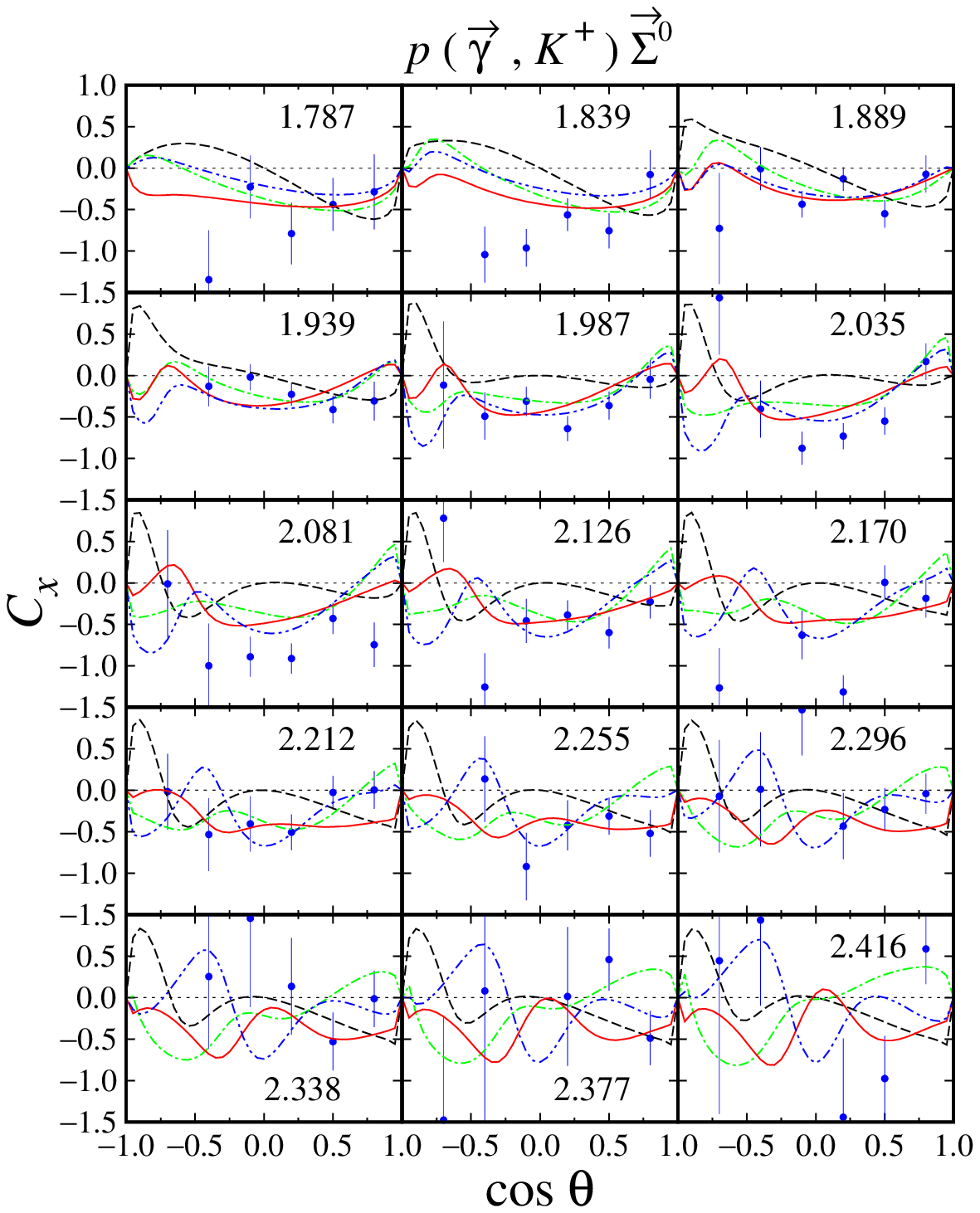}
\caption{As in Fig.~\ref{fig:cx3w}, but for angular distribution.}
\label{fig:cx3th}
\end{figure}

\begin{figure}
\includegraphics[scale=0.8]{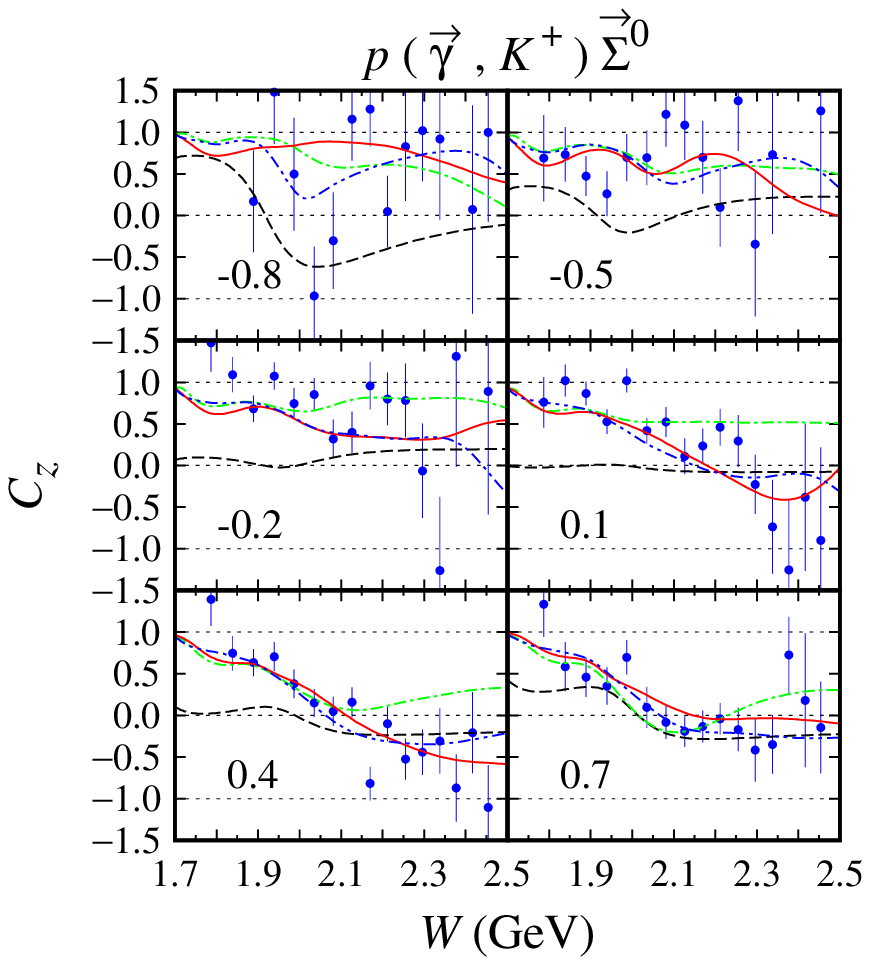}
\caption{As in Fig.~\ref{fig:cx3w}, but for the energy distribution of beam-recoil polarization $C_z$.}
\label{fig:cz3w}
\end{figure}

\begin{figure}
\includegraphics[scale=0.73]{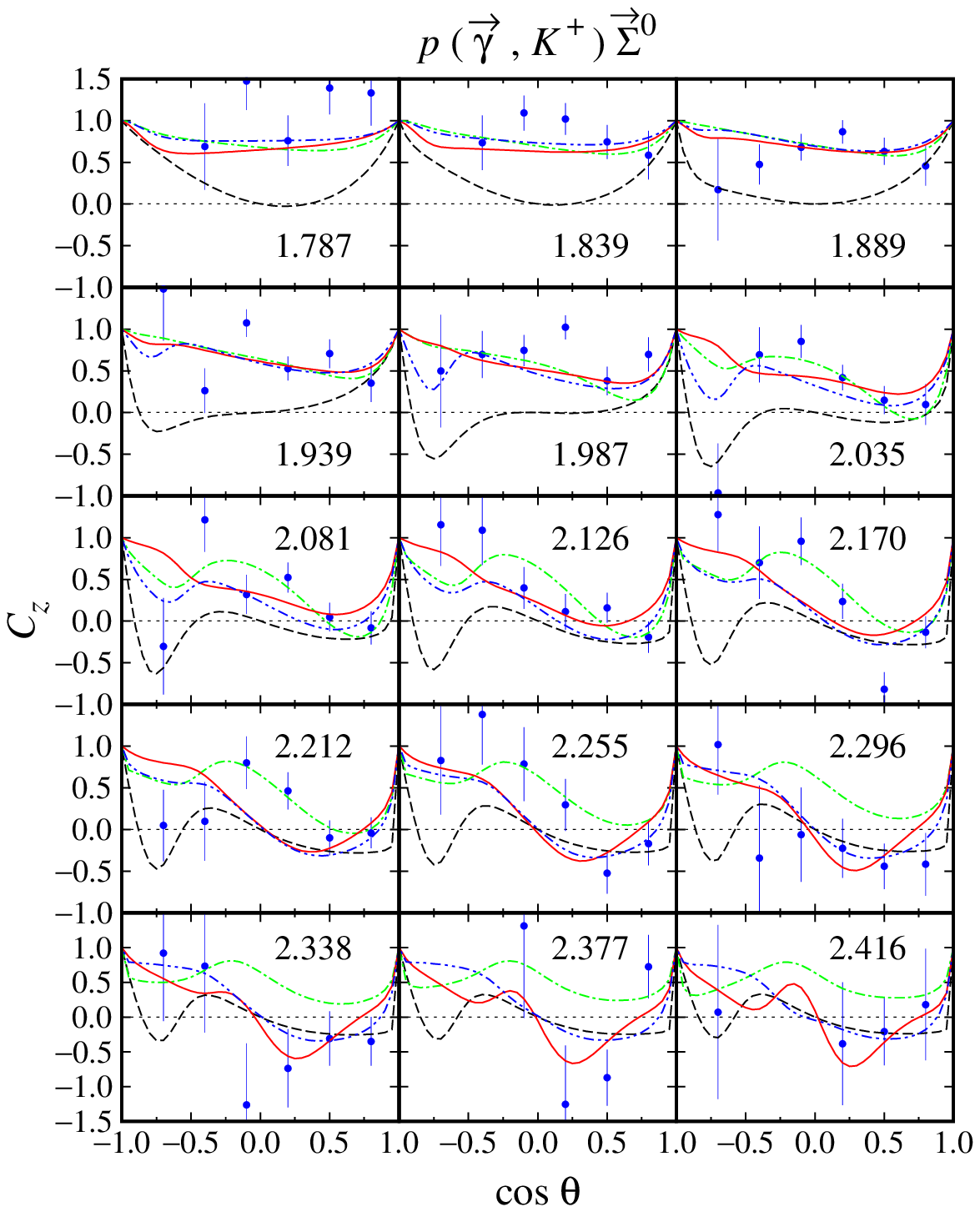}
\caption{As in Fig.~\ref{fig:cz3w}, but for angular distribution.}
\label{fig:cz3th}
\end{figure}

For the $K^+\Sigma^0$ channels we notice that out of 12 possible double
polarization observables, experimental data are only available for four 
types of the beam-recoil observables, i.e., $C_x$, $C_z$, $O_x$, and $O_z$. 
In the case of the $C_x$ and $C_z$, unfortunately, the currently available
experimental data have large error bars, especially in the extreme 
kinematics as shown in Figs.~\ref{fig:cx3w}-\ref{fig:cz3th}. 
As a consequence, all models investigated in the present work yield 
a large variance in this kinematics. As discussed before, the effect
propagates to the calculated cross sections, for which large variation 
can be observed in the extreme kinematics (see Figs.~\ref{fig:difcs3w} 
and \ref{fig:difcs3th}). Furthermore, we also observe this effect in 
the recoil polarization as clearly shown in 
Figs.~\ref{fig:recpo3w} and \ref{fig:recpo3th}.

\begin{figure}
\includegraphics[scale=0.73]{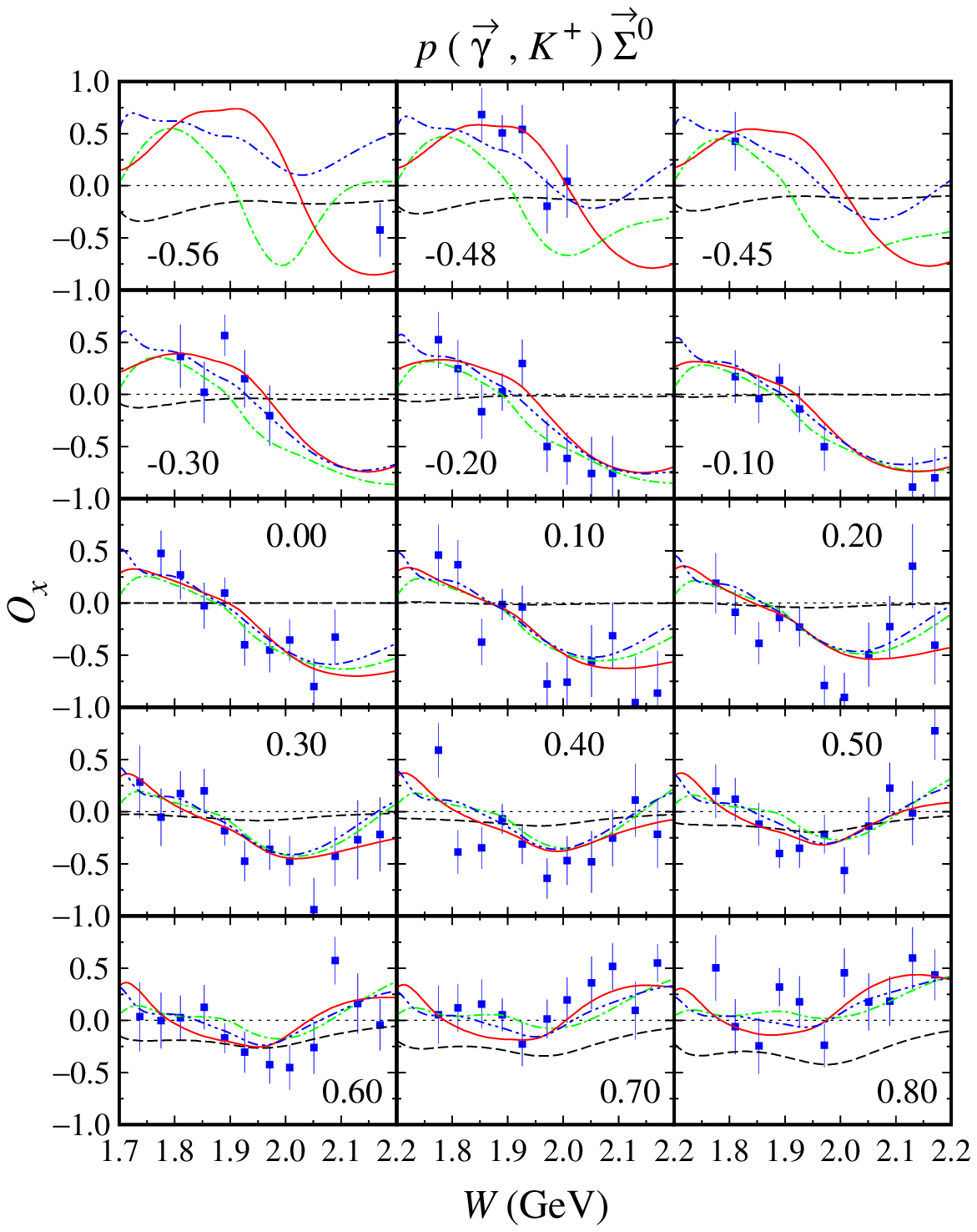}
\caption{As in Fig.~\ref{fig:tarpo3w}, but for the energy distribution of the beam-recoil polarization $O_x$.}
\label{fig:ox3w}
\end{figure}

\begin{figure}
\includegraphics[scale=0.73]{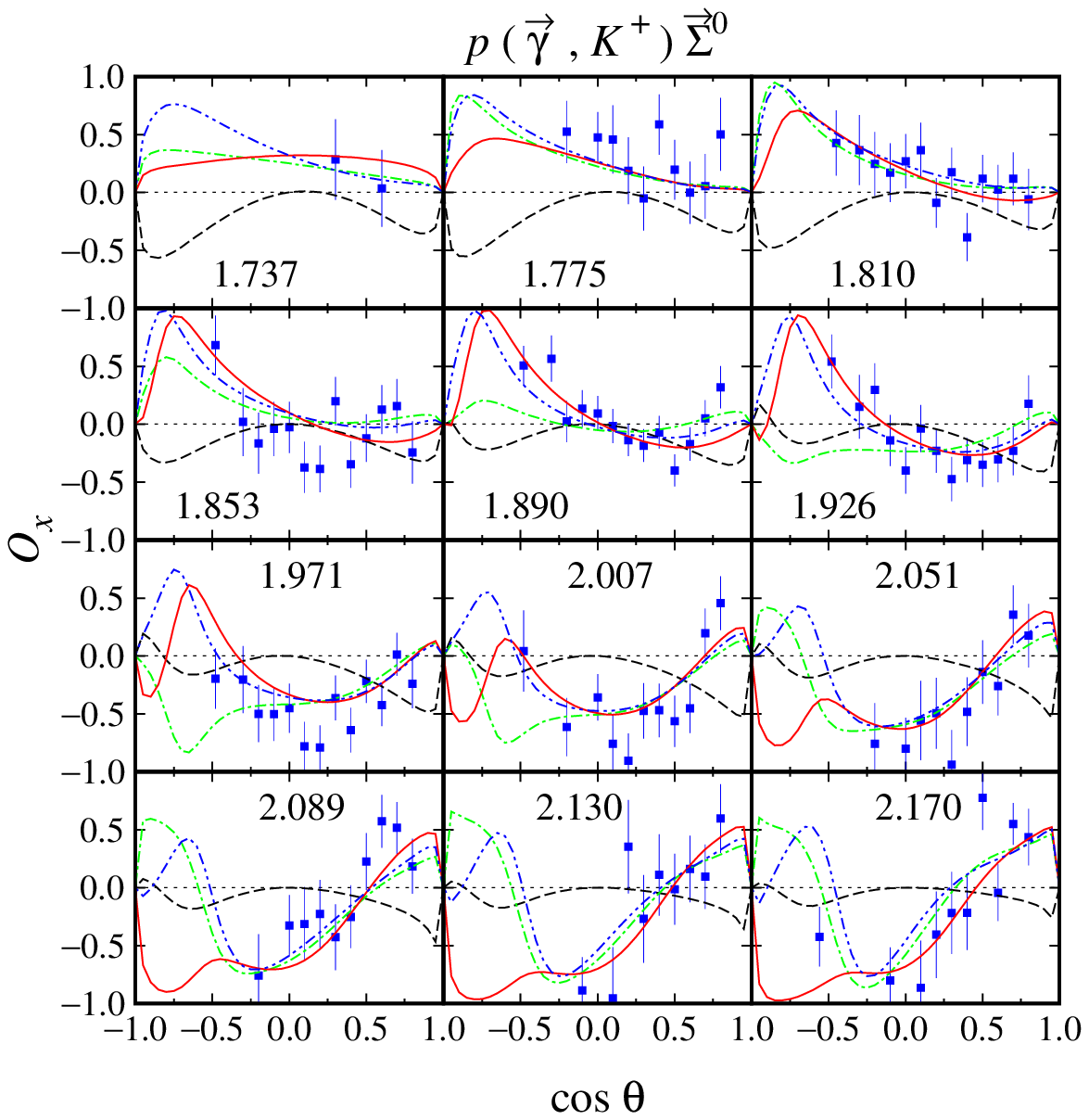}
\caption{As in Fig.~\ref{fig:ox3w}, but for angular distribution.}
\label{fig:ox3th}
\end{figure}

\begin{figure}
\includegraphics[scale=0.73]{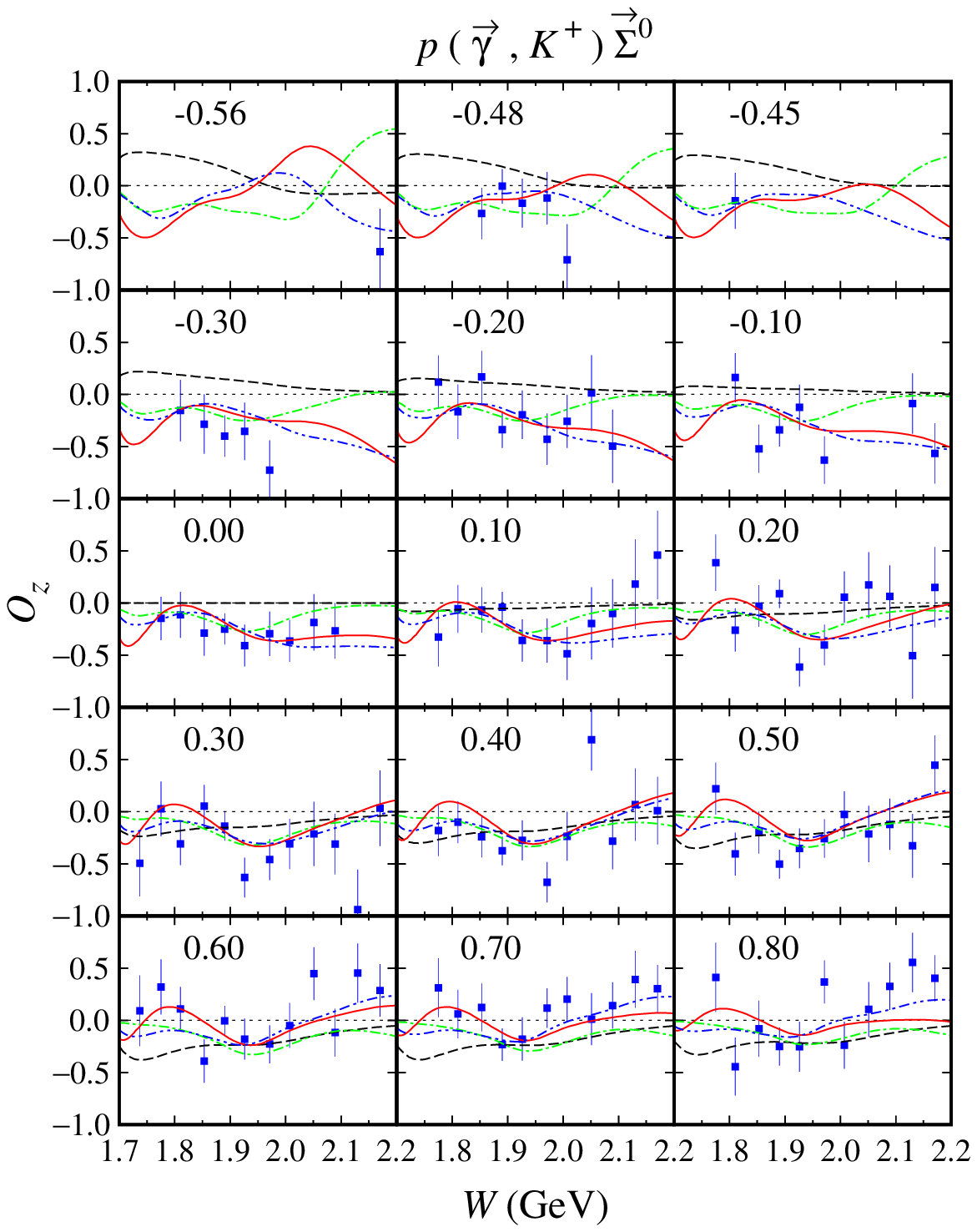}
\caption{As in Fig.~\ref{fig:tarpo3w} but for the energy distribution of the beam-recoil polarization $O_z$.}
\label{fig:oz3w}
\end{figure}

\begin{figure}
\includegraphics[scale=0.73]{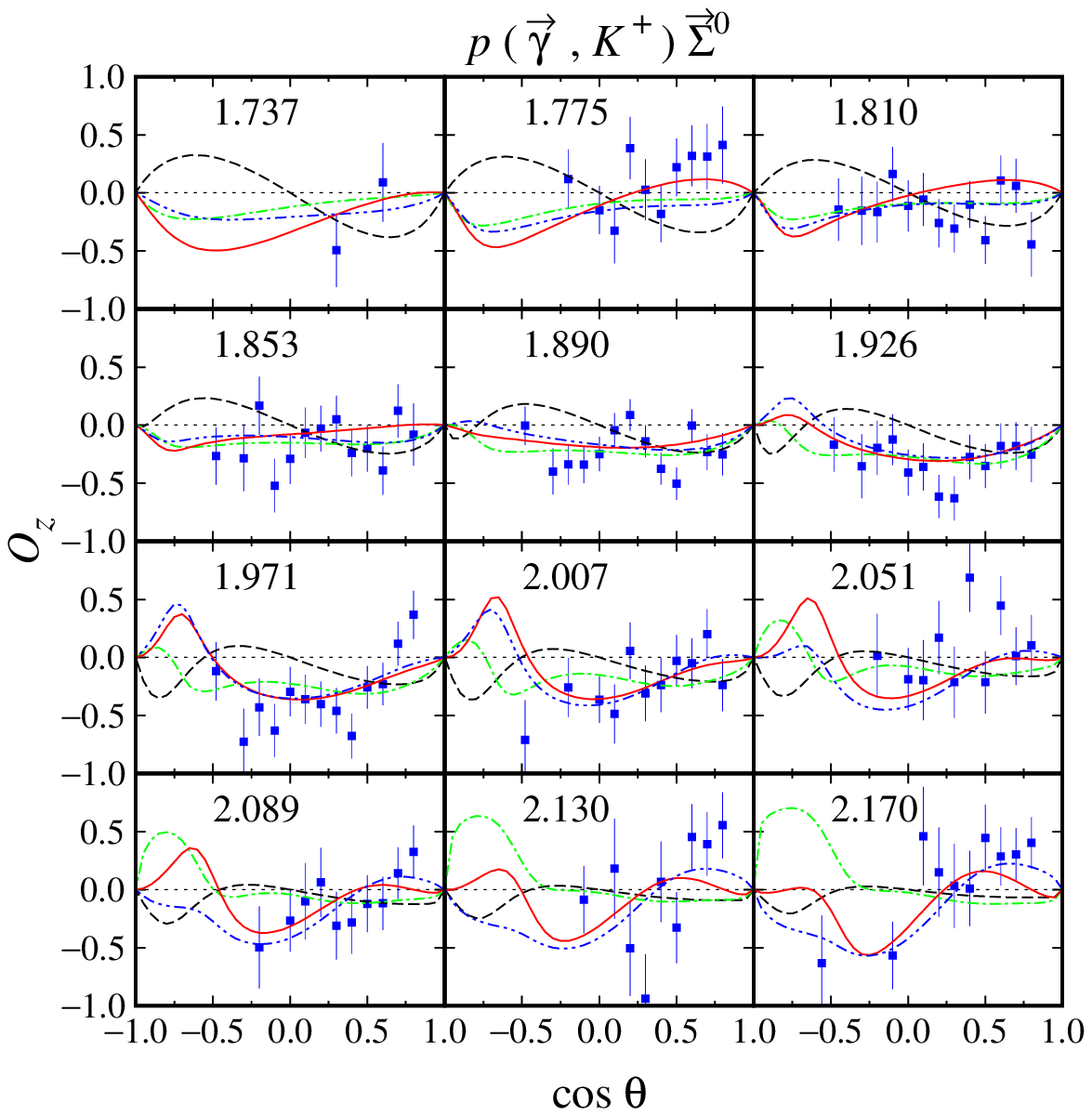}
\caption{As in Fig.~\ref{fig:oz3w}, but for angular distribution.}
\label{fig:oz3th}
\end{figure}

Unlike the double-polarization observables $C_x$ and $C_z$, the available 
data for $O_x$ and $O_z$ are relatively more accurate, as shown in 
Figs.~\ref{fig:ox3w}-\ref{fig:oz3th}. We found that the three models have 
relatively similar trend in the region where experimental data are 
sufficiently available. However, a different situation is shown in the extreme 
kinematics, where experimental data are extremely scarce. Actually, the effects of this 
polarization is small only at the extreme kinematics, where the cross 
section data are accidentally scarce. Nevertheless, to obtain a more 
accurate model, including the background part, these data are urgently 
required, since contribution of the background part in this kinematic is 
stronger than in any other kinematical regions. 



\begin{figure}
\includegraphics[scale=0.73]{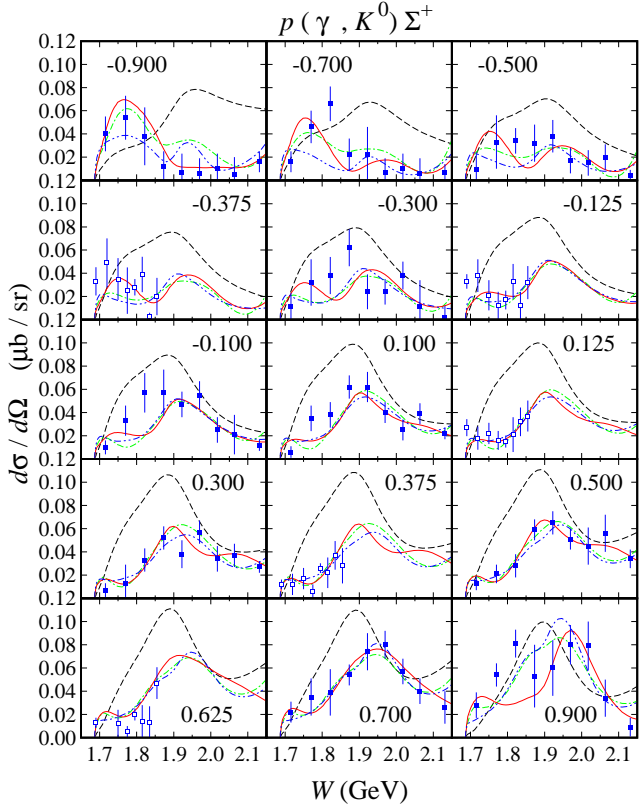}
\caption{Energy distribution of the $\gamma p \to K^0\Sigma^+$ 
  differential cross section obtained from all models. Notation of 
  the curves is as in Fig.~\ref{fig:kstot}. Experimental data are 
  obtained from the SAPHIR 2006 (solid squares~\cite{Lawall:2005np}) 
  and MAMI 2018 (open squares~\cite{Akondi:2018shh}) collaborations.}
\label{fig:difcs4w}
\end{figure}

\begin{figure}
\includegraphics[scale=0.73]{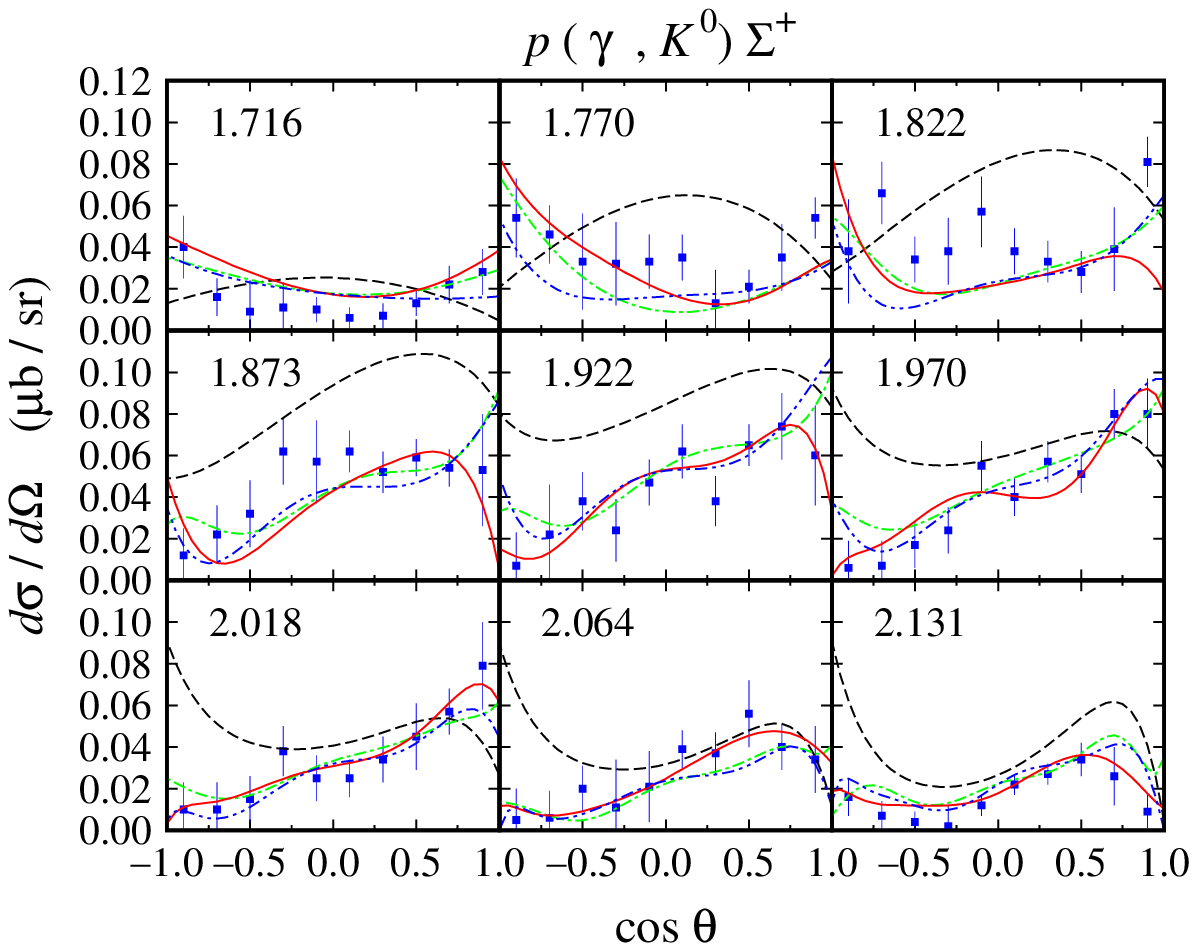}
\caption{As in Fig.~\ref{fig:difcs4w}, but for angular distribution.}
\label{fig:difcs4th}
\end{figure}

\begin{figure*}
\includegraphics[scale=0.85]{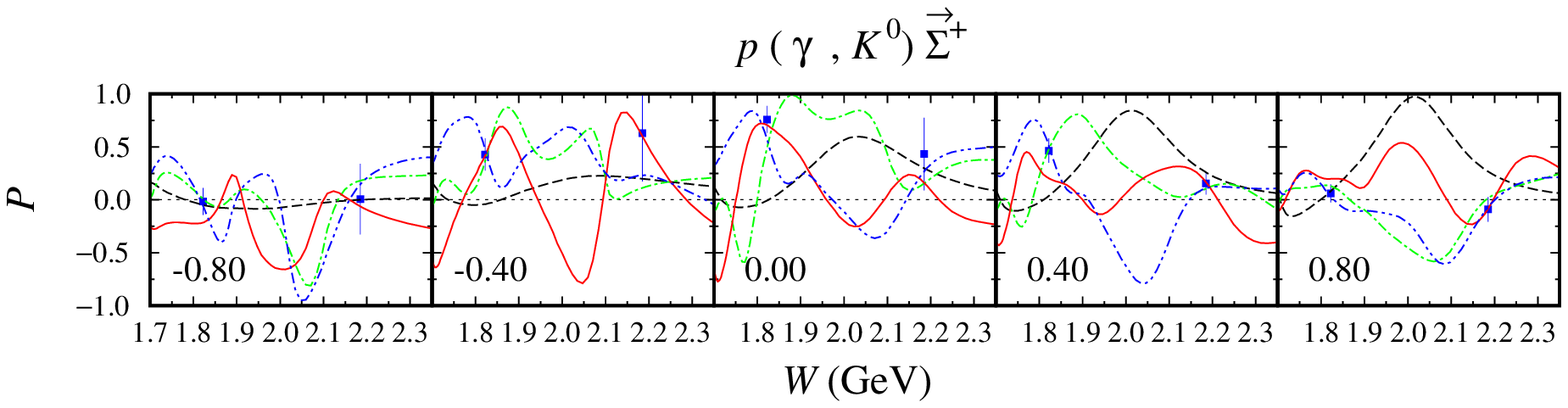}
\caption{As in Fig.~\ref{fig:difcs4w} but for the energy distribution of recoil polarization.}
\label{fig:recpo4w}
\end{figure*}

\begin{figure}[!]
\includegraphics[scale=0.73]{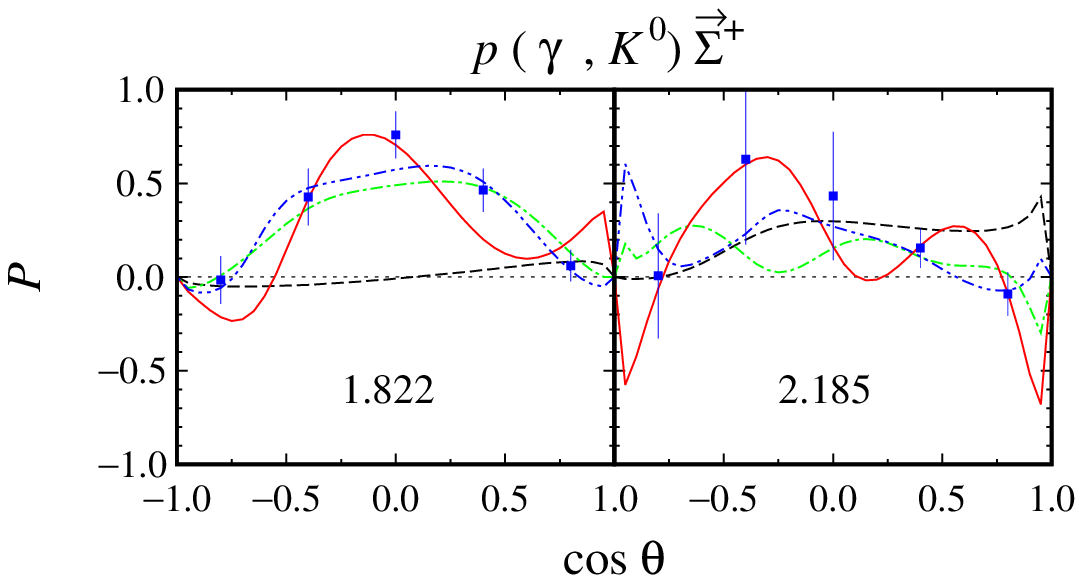}
\caption{As in Fig.~\ref{fig:recpo4w}, but for angular distribution.}
\label{fig:recpo4th}
\end{figure}

\begin{figure}
\includegraphics[scale=0.73]{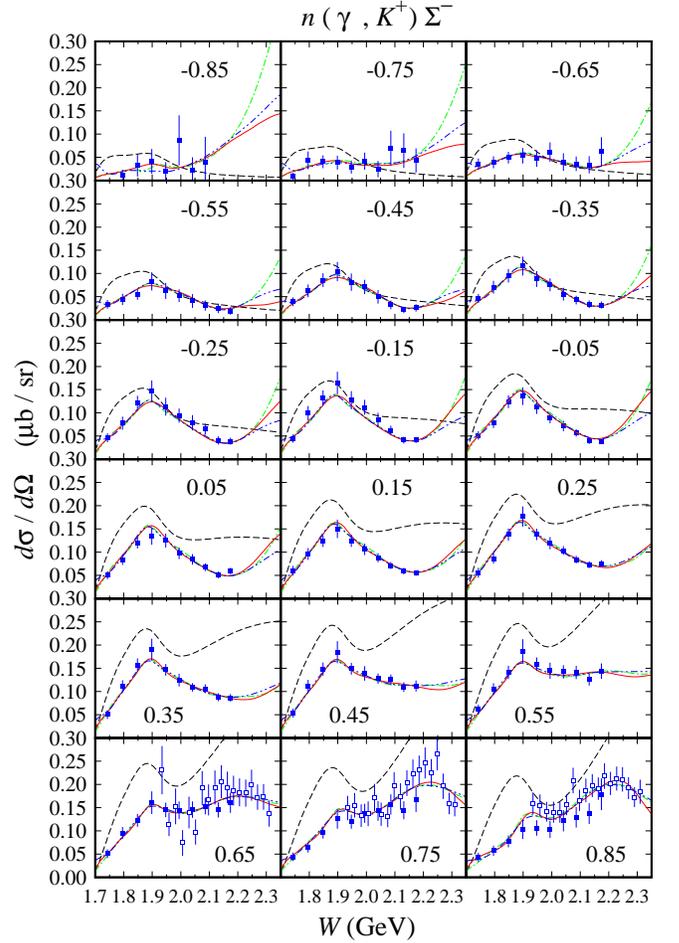}
\caption{Energy distribution of the  $\gamma n \to K^+\Sigma^-$ differential 
  cross section obtained from all models. Notation of the curves is as in 
  Fig.~\ref{fig:kstot}. Experimental data are obtained from LEPS 2006 
  (open squares~\cite{Kohri:2006yx}) and CLAS 2010 (solid squares~\cite{AnefalosPereira:2009zw}) 
  collaborations.}
\label{fig:difcs5w}
\end{figure}

\begin{figure}
\includegraphics[scale=0.73]{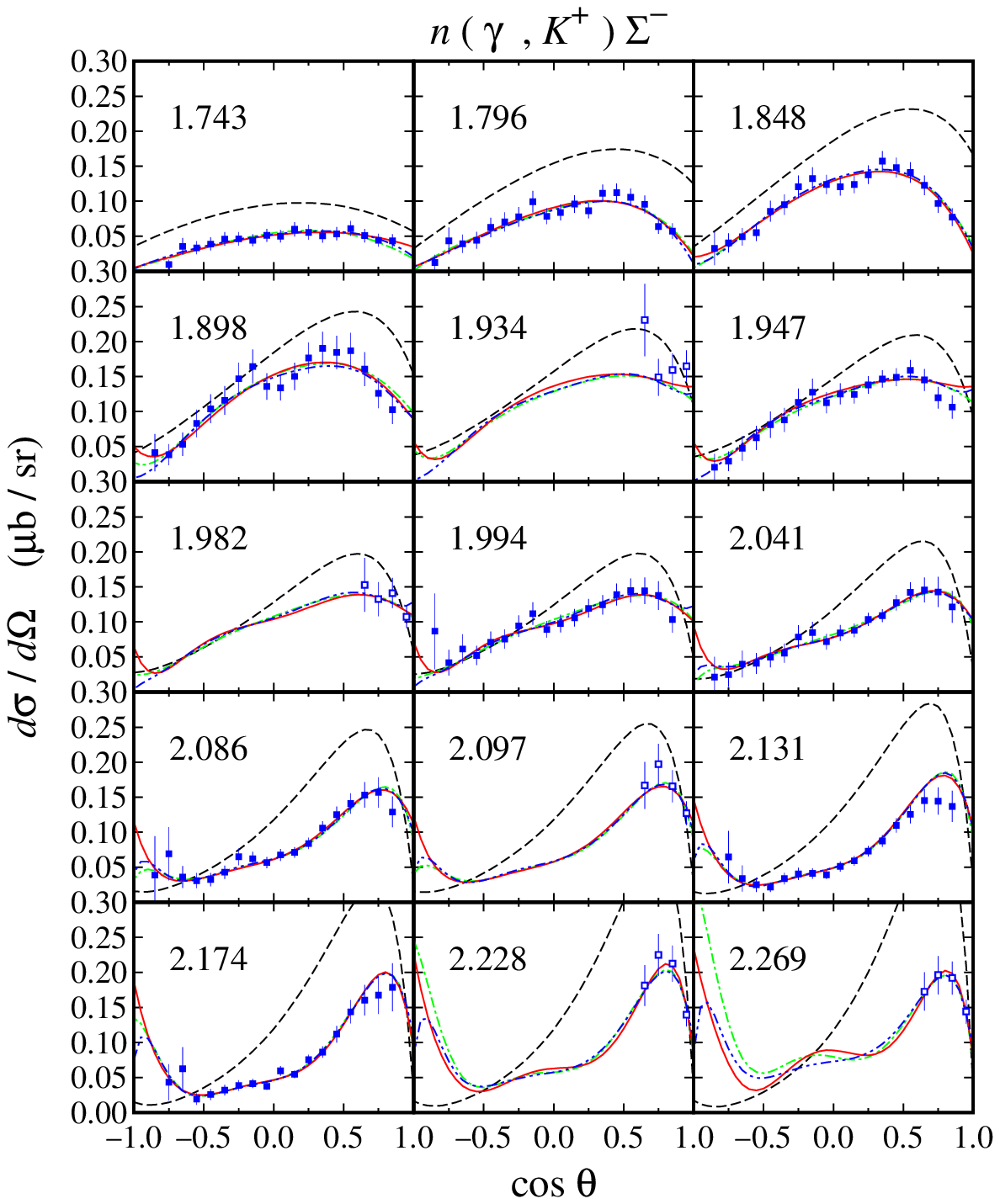}
\caption{As in Fig.~\ref{fig:difcs5w}, but for angular distribution.}
\label{fig:difcs5th}
\end{figure}

\begin{figure}
\includegraphics[scale=0.80]{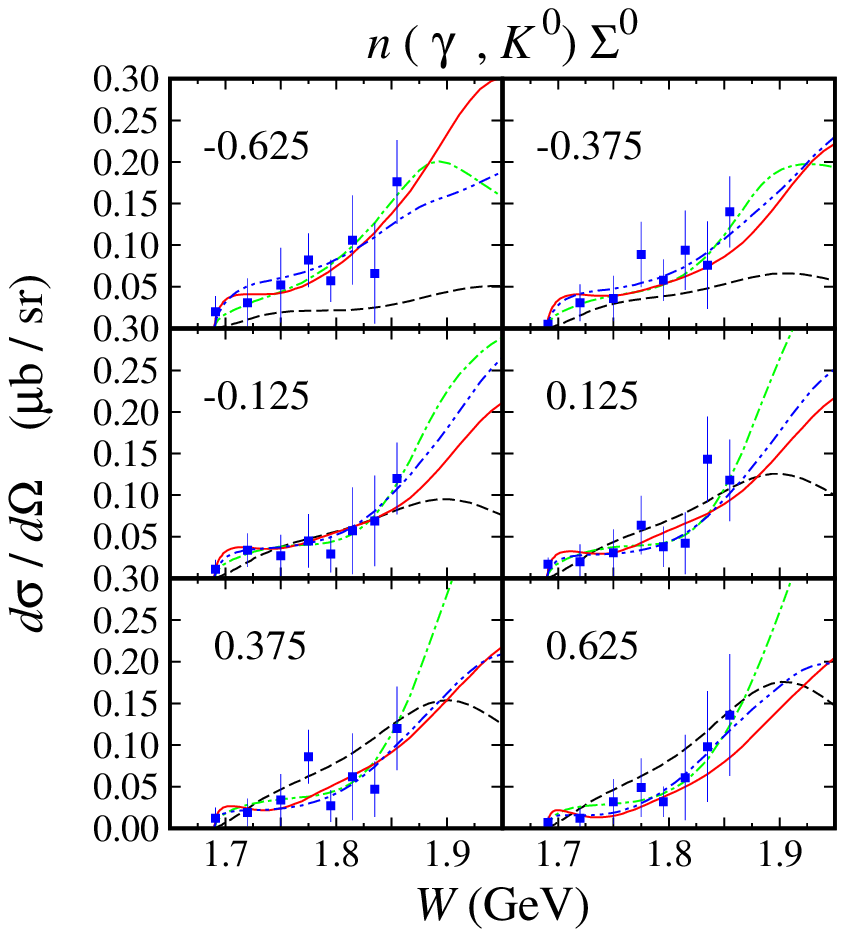}
\caption{Energy distribution of the  $\gamma n \to K^0\Sigma^0$ differential cross 
  section obtained from all models. Notation of the curves is as in Fig.~\ref{fig:kstot}. 
  Experimental data are obtained from MAMI 2018 (solid squares~\cite{Akondi:2018shh}) 
  collaborations.}
\label{fig:difcs6w}
\end{figure}
\begin{figure}
\includegraphics[scale=0.80]{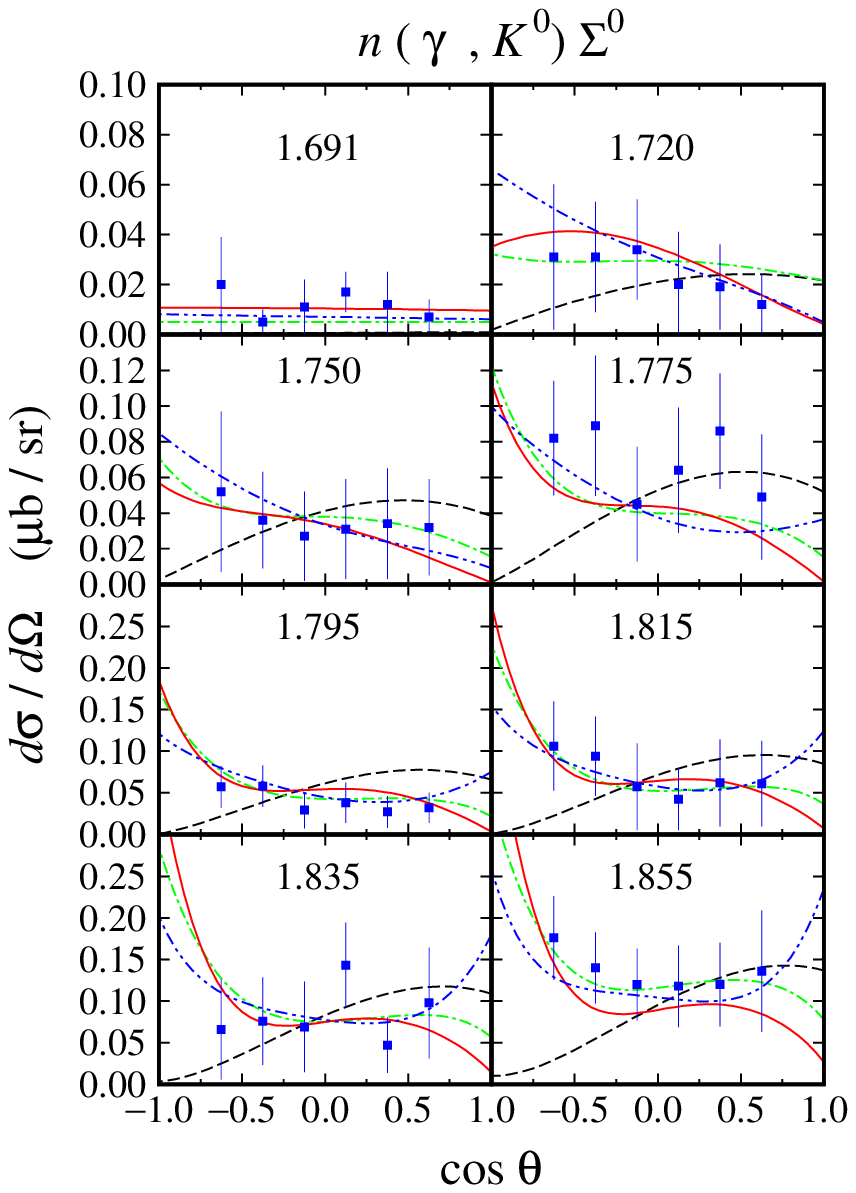}
\caption{As in Fig.~\ref{fig:difcs6w}, but for angular distribution.}
\label{fig:difcs6th}
\end{figure}

\begin{figure}
\includegraphics[scale=0.85]{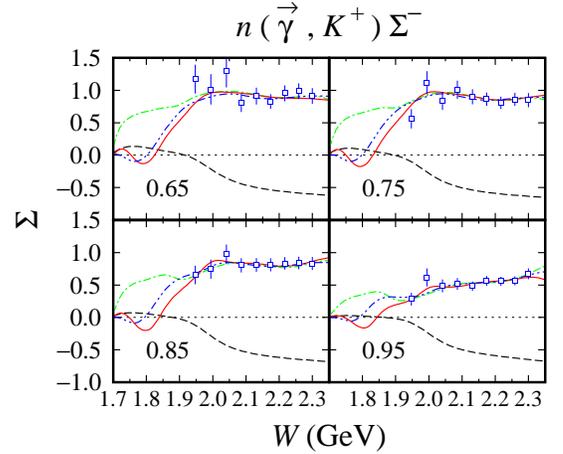}
\caption{As in Fig.~\ref{fig:difcs5w}, but for the energy distribution of photon asymmetry.}
\label{fig:phas5w}
\end{figure}
\begin{figure}
\includegraphics[scale=0.85]{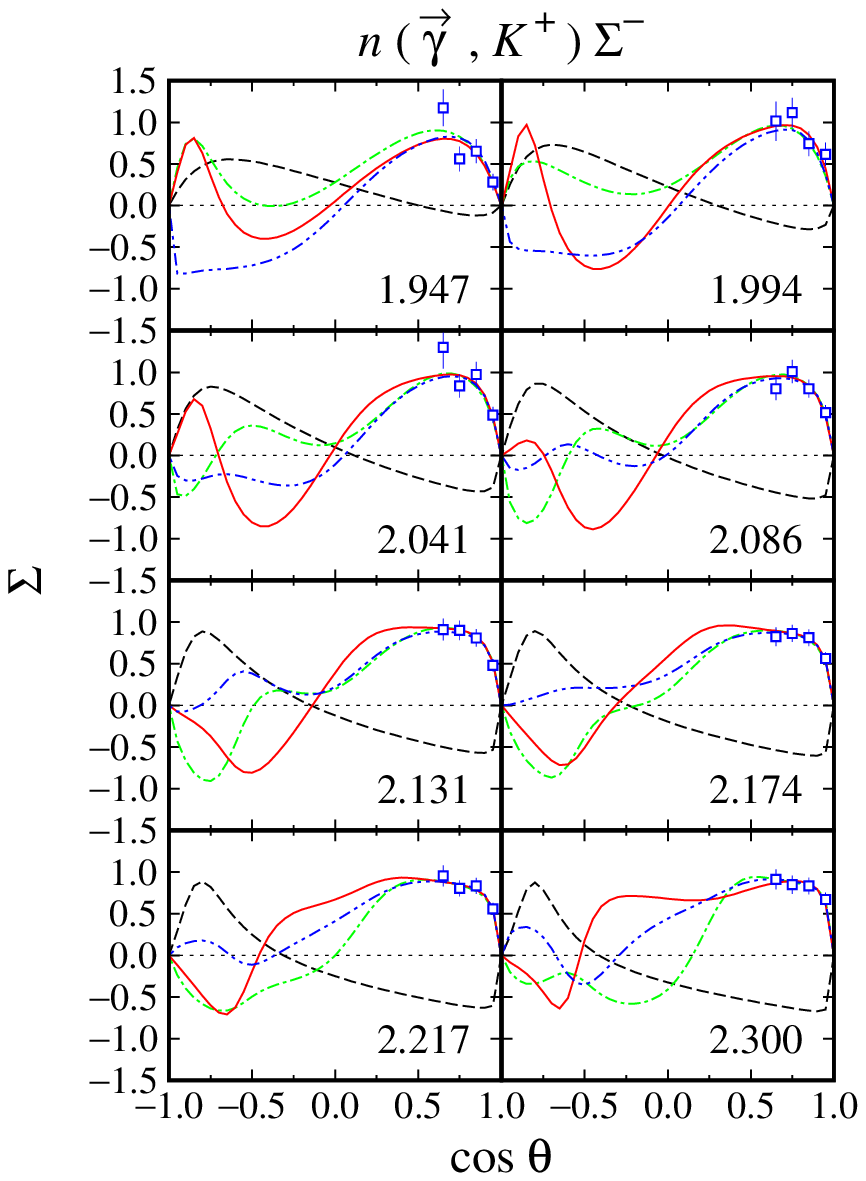}
\caption{As in Fig.~\ref{fig:phas5w}, but for angular distribution.}
\label{fig:phas5th}
\end{figure}

As stated before, the model HFF-P3 yields the best agreement with 
experimental data,  although the difference is not large compared to 
the model HFF-G in the case of $\gamma p \to K^0\Sigma^+$ channel. 
From Figs.~\ref{fig:difcs4w} and \ref{fig:difcs4th}, we may conclude that 
contribution to the first peak in the differential cross section 
at $W\approx 1.75$ GeV (see Fig.~\ref{fig:difcs4w}) stems mostly from 
the backward angles, since as shown in Fig.~\ref{fig:difcs4th} the
cross section in this energy region is backward peaking. The angular 
distributions of differential cross section obtained from the model HFF-P3 
exhibit a clearer picture of the origin of both peaks in the total cross 
section. By comparing with the results from the other two models we might conclude that the 
difference between them originates from different ways of reproducing 
the recoil polarization data shown in Figs.~\ref{fig:recpo4w}-\ref{fig:recpo4th}, 
since, as shown in Figs.~\ref{fig:difcs4w} and \ref{fig:difcs4th}, both SAPHIR 
and MAMI data do not show significant discrepancy. 

It is also obvious that at very forward and backward angles, only the model HFF-P3 
shows different recoil polarization. The difference originates from the lack 
of data in the extreme kinematics from the MAMI collaboration. Thus, we might 
conclude that the different behavior shown by all models originates from 
the freedom to predict the missing data in the extreme kinematics.

The other $\Sigma$ channels that use neutron as target exhibit a similar pattern 
as shown in Figs.~\ref{fig:difcs5w}-\ref{fig:difcs6th}. As expected, there are 
differences in the kinematical regions, where experimental data are lacking. Interestingly, 
we observe a unique behavior in the prediction of the model HFF-G for the 
$\gamma n \to K^+\Sigma^-$, i.e., there are tiny peaks near the threshold region 
at backward and forward angles. The peaks, which will be discussed later 
when we discuss the resonances properties, originate from the contribution of 
high spin resonances which only appears near threshold in the model HFF-G. 
Unfortunately, this behavior cannot be further explored due to the lack of 
experimental data in this kinematics.  As shown in Figs.~\ref{fig:phas5w} and
\ref{fig:phas5th}, the same situation also happens in the case of the polarization 
asymmetry $\Sigma$ that usually provides a severe constraint 
to the model. Nevertheless, we will discuss this topic from another perspective 
in the following subsection.

\subsection{Resonances Properties}
Having constructed an isobar model that can nicely reproduce all 
$K\Sigma$ photoproduction data we are in the position to study the 
properties of baryon resonances through their electromagnetic and
hadronic interactions. The resonance properties that are of interest 
in the hadronic physics community are the resonance mass and width 
evaluated at its pole position as well as the extracted resonance partial 
decay width. Evaluation of the resonance mass and width at pole 
position has a clear advantage, since the result is model independent. 
Therefore, the properties of resonances evaluated at pole position 
in the present study are comparable with those obtained from other
model-independent investigations. 

The result is shown in Table~\ref{tab:pole}, where we compare the 
resonance masses and widths evaluated at their poles obtained from the 
present work with those given by PDG. For completeness, we present the
corresponding Breit-Wigner masses and widths in Table~\ref{tab:mwbw}
of Appendix~\ref{app:mwbw}. The calculated resonance masses 
show a good agreement with the PDG values, especially in the case of
model HFF-G. On the other hand, the calculated widths show some 
discrepancies with the PDG ones. There is no obvious pattern observed
in these discrepancies and, in fact, it seems to be random for all models. 
We believe that this phenomenon originates from the use of single channel analysis, 
where the resonance widths are not unitary defined. In the case of 
the current best models (HFF-G and HFF-P3), there are two adjacent 
resonances that show an interesting phenomenon, i.e., the $N(1700)D_{13}$ 
and $N(1720)P_{13}$ states. In  these models the resonances produce different 
cross section behavior near the threshold by switching their pole 
positions. However, in the model HFF-G the pole positions of these 
resonances are closer to the PDG values. Put in other words, compared
to other nucleon resonances the $N(1700)D_{13}$ state is very likely to be 
important in the threshold region. Nevertheless, this finding still needs 
further investigation once the $K\Sigma$ photoproduction data near 
threshold are sufficiently available.
\begin{table*}
\centering
\caption{Masses and widths of the nucleon and $\Delta$ resonances evaluated 
  at their pole positions from the present work and PDG \cite{pdg}.}
\label{tab:pole}
\begin{ruledtabular}
\begin{tabular}{cccccccccc}
 Resonances & $J^P$ & \multicolumn{2}{c}{PDG} & \multicolumn{2}{c}{HFF-P1} & \multicolumn{2}{c}{HFF-P3} & \multicolumn{2}{c}{HFF-G}\\
 \cline{3-4}  \cline{5-6}\cline{7-8}\cline{9-10}
  && $M_p$ (MeV) & $\Gamma_p$ (MeV) & $M_p$ (MeV) & $\Gamma_p$ (MeV) & $M_p$ (MeV) & $\Gamma_p$ (MeV) & $M_p$ (MeV) & $\Gamma_p$ (MeV) \\
 \hline
 $N(1440)P_{11}$ & $1/2^{+}$ & $1370\pm10$ & $175\pm15$ & 1324 & 188 & 1351 & 206 & 1398 & 174 \\
 $N(1520)D_{13}$ & $3/2^{-}$ & $1510\pm 5$ & $110^{+10}_{-5}$ & 1489 & 100 & 1489 & 100 & 1505 &  86 \\
 $N(1535)S_{11}$ & $1/2^{-}$ & $1510\pm10$ & $130\pm20$ & 1530 & 129 & 1530 & 129 & 1515 & 185 \\
 $N(1650)S_{11}$ & $1/2^{-}$ & $1655\pm15$ & $135\pm35$ & 1664 & 176 & 1658 & 121 & 1644 & 112 \\
 $N(1675)D_{15}$ & $5/2^{-}$ & $1660\pm 5$ & $135^{+15}_{-10}$ & 1643 & 136 & 1651 & 114 & 1653 & 137 \\
 $N(1680)F_{15}$ & $5/2^{+}$ & $1675^{+5}_{-10}$ & $120^{+15}_{-10}$ & 1667 &  98 & 1655 & 108 & 1668 &  98 \\
 $N(1700)D_{13}$ & $3/2^{-}$ & $1700\pm50$ & $200\pm100$ & 1630 & 111 & 1635 &  97 & 1718 & 158 \\
 $N(1710)P_{11}$ & $1/2^{+}$ & $1700\pm20$ & $120\pm 40$ & 1705 &  47 & 1679 &  41 & 1685 &   3 \\
 $N(1720)P_{13}$ & $3/2^{+}$ & $1675\pm15$ & $250^{+150}_{-100}$ & 1665 & 300 & 1712 & 222 & 1625 & 253 \\
 $N(1860)F_{15}$ & $5/2^{+}$ & $1830^{+120}_{-60}$ & $250^{+150}_{-50}$ & 1787 & 156 & 1773 & 162 & 1778 & 155 \\
 $N(1875)D_{13}$ & $3/2^{-}$ & $1900\pm50$ & $160\pm60$ & 1757 & 219 & 1757 & 219 & 1757 & 219 \\
 $N(1880)P_{11}$ & $1/2^{+}$ & $1860\pm40$ & $230\pm50$ & 1831 & 166 & 1887 & 204 & 1895 & 173 \\
 $N(1895)S_{11}$ & $1/2^{-}$ & $1910\pm20$ & $110\pm30$ & 1893 &  90 & 1893 & 124 & 1893 &  90 \\
 $N(1900)P_{13}$ & $3/2^{+}$ & $1920\pm20$ & $150\pm50$ & 1899 & 239 & 1918 & 148 & 1833 & 228 \\
 $N(1990)F_{17}$ & $7/2^{+}$ & $2030\pm65$ & $240\pm60$ & 2044 & 273 & 1978 & 167 & 1999 & 169 \\
 $N(2000)F_{15}$ & $5/2^{+}$ & $2030\pm40$ & $380\pm60$ & 1978 & 232 & 1963 & 229 & 1963 & 229 \\
 $N(2060)D_{15}$ & $5/2^{-}$ & $2070^{+60}_{-50}$ & $400^{+30}_{-50}$ & 1968 & 334 & 1937 & 322 & 1970 & 332 \\
 $N(2120)D_{13}$ & $3/2^{-}$ & $2100\pm50$ & $280\pm60$ & 2029 & 274 & 2031 & 274 & 2125 & 287 \\
 $N(2190)G_{17}$ & $7/2^{-}$ & $2100\pm50$ & $400\pm100$ & 2142 & 211 & 2142 & 211 & 2046 & 196 \\
 $N(2220)H_{19}$ & $9/2^{+}$ & $2170^{+30}_{-40}$ & $400^{+80}_{-40}$ & 2131 & 202 & 2130 & 197 & 2130 & 197 \\
 $N(2290)G_{19}$ & $9/2^{-}$ & $2200\pm50$ & $420^{+80}_{-50}$ & 2193 & 219 & 2265 & 229 & 2193 & 219 \\
 $\Delta(1232)P_{33}$ & $3/2^{+}$ & $1210\pm1$ & $100\pm2$ & 1205 &  82 & 1211 &  81 & 1210 &  83 \\
 $\Delta(1600)P_{33}$ & $3/2^{+}$ & $1510\pm50$ & $270\pm70$ & 1457 & 168 & 1454 & 171 & 1595 & 313 \\
 $\Delta(1620)S_{31}$ & $1/2^{-}$ & $1600\pm10$ & $120\pm20$ & 1598 & 152 & 1626 & 152 & 1659 & 152 \\
 $\Delta(1700)D_{33}$ & $3/2^{-}$ & $1665\pm25$ & $250\pm50$ & 1646 & 161 & 1602 & 211 & 1704 & 186 \\
 $\Delta(1900)S_{31}$ & $1/2^{-}$ & $1865\pm35$ & $240\pm60$ & 1938 & 330 & 1938 & 330 & 1938 & 330 \\
 $\Delta(1905)F_{35}$ & $5/2^{+}$ & $1800\pm30$ & $300\pm40$ & 1797 & 212 & 1801 & 177 & 1847 & 185 \\
 $\Delta(1910)P_{31}$ & $1/2^{+}$ & $1860\pm30$ & $300\pm100$ & 1859 & 317 & 1875 & 269 & 1864 & 304 \\
 $\Delta(1920)P_{33}$ & $3/2^{+}$ & $1900\pm50$ & $300\pm100$ & 1893 & 193 & 1873 & 290 & 1913 & 293 \\
 $\Delta(1930)D_{35}$ & $5/2^{-}$ & $1880\pm40$ & $280\pm50$ & 1933 & 199 & 1971 & 205 & 1971 & 203 \\
 $\Delta(1940)D_{33}$ & $3/2^{-}$ & $1950\pm100$ & $350\pm150$ & 1880 & 349 & 1840 & 332 & 1864 & 284 \\
 $\Delta(1950)F_{37}$ & $7/2^{+}$ & $1880\pm10$ & $240\pm20$ & 1848 & 208 & 1907 & 179 & 1872 & 174 \\
 $\Delta(2000)F_{35}$ & $5/2^{+}$ & $2150\pm100$ & $350\pm100$ & 2081 & 328 & 2196 & 237 & 2216 & 374 \\
 $\Delta(2300)H_{39}$ & $9/2^{+}$ & $2370\pm80$ & $420\pm160$ & 2341 & 200 & 2378 & 204 & 2354 & 201 \\
 $\Delta(2400)G_{39}$ & $9/2^{-}$ & $2260\pm60$ & $320\pm160$ & 2409 & 329 & 2409 & 329 & 2704 & 402 \\
\end{tabular}
\end{ruledtabular}
\end{table*}

\begin{figure}
\includegraphics[scale=0.22,angle=-90,width=7cm]{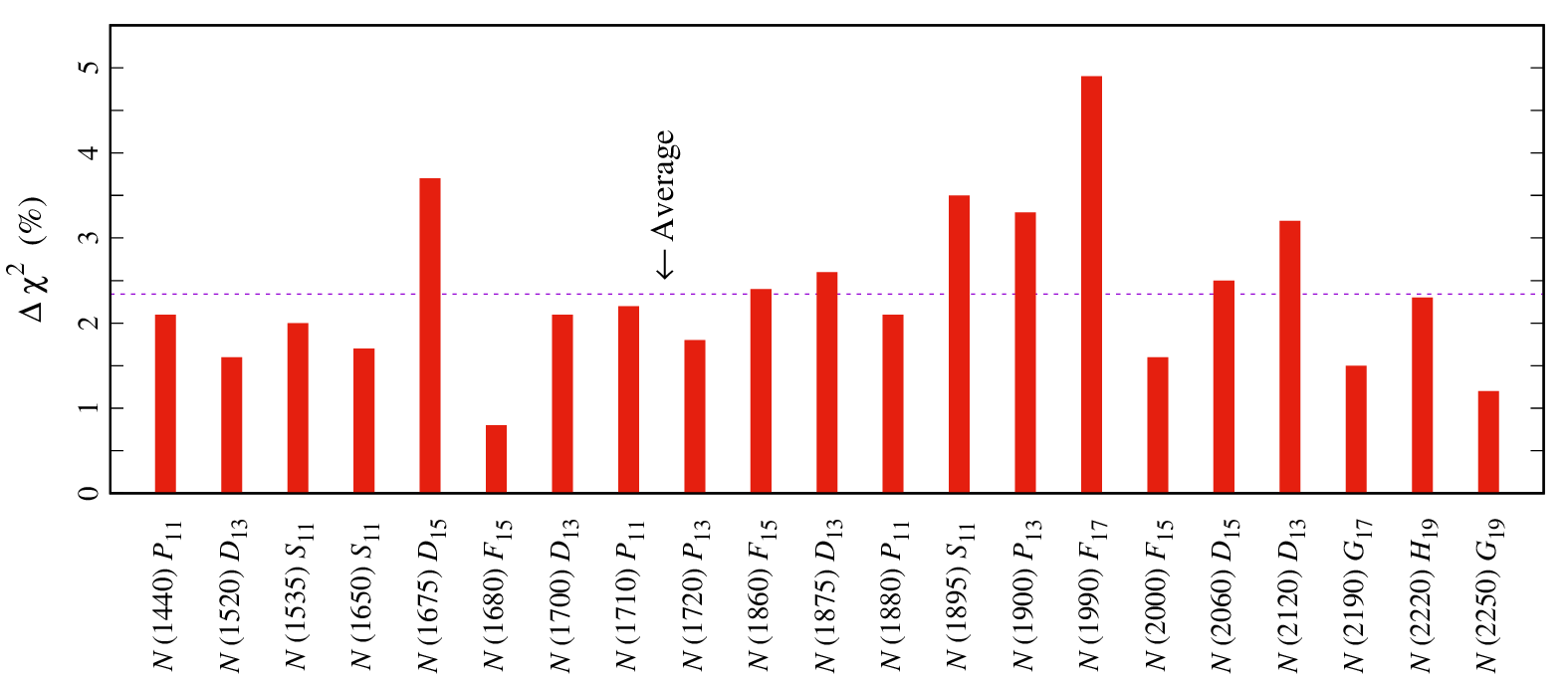}
\caption{The significance of nucleon resonances in the $K\Sigma$ 
  photoproduction obtained from the model HFF-P3.
  The numerical values are calculated from Eq.~(\ref{eq:par_res}) and
  their average is indicated by the dashed line.}
\label{fig:nstar}
\end{figure}

\begin{figure}
\includegraphics[scale=0.22,angle=-90,width=7cm]{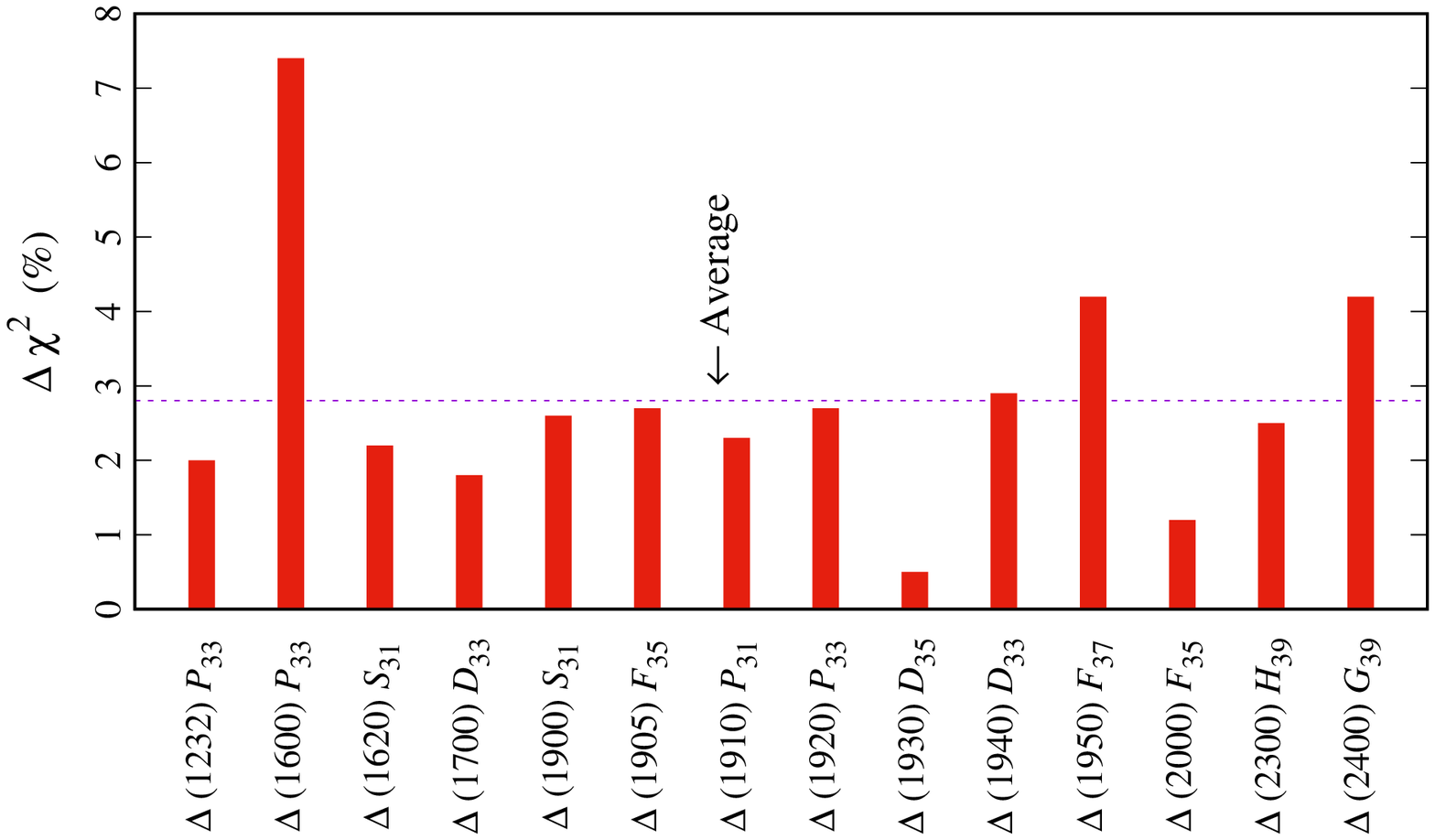}
\caption{As in Fig.~\ref{fig:nstar}, but for the $\Delta$ resonances.}
\label{fig:dstar}
\end{figure}

It is also interesting to explore how the resonances complement to each other 
through their significance in the $K\Sigma$ photoproduction reaction. In 
Figs.~\ref{fig:nstar}-\ref{fig:dstar} we show the significance of 
each resonance for the best model, i.e., model HFF-P3. 
We observe that in average the contribution of each resonance is
relatively small, i.e., under $8\%$. This is understandable 
since the number of resonances used in the model is large. As a consequence, the task to
produce different structures in all calculated observables could
be easily distributed to all resonances in the model. It is 
therefore obvious that the role 
of the excluded resonance will be immediately replaced by the 
adjacent one. Of course, there are a number of resonances 
that exhibit a relatively stronger or weaker significance compared 
to the other ones. For the stronger one, there is a resonance that 
contributes to the background because its mass is below the
threshold energy, i.e., the $\Delta(1600)P_{33}$ state with $J^P = 3/2^+$. 
Apparently, the adjacent resonances cannot replace this resonance. 
As shown in Fig.~\ref{fig:dstar}, this resonance is the most important 
$\Delta$ resonance in the present work. 

From Table \ref{tab:pole} we might expect that both $N(1990)F_{17}$ and 
$\Delta(1950)F_{37}$ resonances could complement to each other. 
However, Figs.~\ref{fig:nstar}-\ref{fig:dstar} indicate that they 
still have strong impact on the agreement between model calculation
and experimental data. This can be understood because there are no 
resonances with spin equals to $7/2$ and adjacent mass. 
Another example that shows the importance of high spin resonances 
is exhibited by the $\Delta(2400)G_{39}$ state. This resonance has 
a relatively high significance compared to the other ones. By looking
at the PDG estimate in Table \ref{tab:pole} we obviously find that 
the extracted masses are significantly heavier. We note that this
occurs in both Breit-Wigner and pole position methods.  
Furthermore, Table \ref{tab:pole}
also indicates that in the model HFF-G the extracted mass is about 500 MeV 
heavier than the PDG estimate. Therefore, we might conclude that 
this resonance is responsible for improving the agreement 
with experimental data in the higher energy region. Nevertheless,
for a more conclusive finding, investigation at this energy region is strongly advocated in the future.
Since we have observed similar cases in the present work, we 
might conclude that the high spin resonances are indispensable  
in the $K\Sigma$ photoproduction process. The role of spin-7/2 and
9/2 resonances in the $K^+\Lambda$ photoproduction has been thoroughly
analyzed in Ref.~\cite{Clymton:2017nvp}. For spin-11/2 and 13/2 nucleon
and delta resonances the analysis is still ongoing and will be published
soon \cite{lala-2020}. 

Figures \ref{fig:nstar} and \ref{fig:dstar} also show the less 
significance resonances, e.g., the $N(1680)F_{15}$ and $\Delta(1930)D_{35}$,
which have $\Delta\chi^2$ less than $1\%$. In the case of the $N(1680)F_{15}$ 
resonance this is understandable, since it contributes to the background 
part of the model and, furthermore, there is a spin $5/2$ resonance 
with adjacent mass, i.e., the $N(1675)D_{15}$. For the $\Delta(1930)D_{35}$ resonance, 
although there is evidence for its branching to the $K\Sigma$
channels, the present work indicates that this state is less significance. 
A closer look to the mass position reveals that the mass of this 
resonance lies among the masses of six $\Delta$ resonances 
from 1900 to 1950 MeV (see Fig. \ref{fig:dstar}). We believe that 
the less significance of the $\Delta(1930)D_{35}$ resonance originates
from this phenomenon. As a consequence, if we needed to simplify the 
model but not its accuracy, this resonance could be excluded. 

In Tables~\ref{tab:brbwproton} and \ref{tab:brbwneutron} we compare 
the resonance partial widths obtained in the present analysis and 
those listed by PDG. Table~\ref{tab:brbwproton} shows the comparison 
for the proton channels, whereas Table \ref{tab:brbwneutron}
shows that for the neutron channels. Since the value of partial width 
for $\Delta$ resonances is the same for both proton and neutron
channels, we only show them in Table~\ref{tab:brbwproton}. 
From Tables~\ref{tab:brbwproton} and \ref{tab:brbwneutron}, 
we may conclude that most of the extracted partial widths are in 
good agreement with those listed by PDG. Nevertheless, we also note
that there are large discrepancies in the case of the $N(1900)P_{13}$,
$N(2060)D_{15}$, and $\Delta(1920)P_{33}$. This would be an interesting
phenomenon for future investigation, since for the three resonances 
the discrepancies between our best model
HFF-G and the PDG values are milder.

In our previous study~\cite{Clymton:2017nvp} we also observed a similar
phenomenon, i.e., the extracted partial widths for a number of resonances
are found to be very large. Presumably, this is caused by the large number 
of resonances involved in the model, which creates strong interferences 
among the resonances and causes irrelevant contribution of certain resonances
at energy far from the resonance masses. In the present work we found 
that this problem can be overcome by using the Gaussian hadronic form 
factor given in the HFF-G model, which strongly suppresses the amplitude
except near the resonance mass. Comparison among the three models 
given in Tables \ref{tab:brbwproton} and \ref{tab:brbwneutron} shows 
that the Gaussian form factor reduces the extracted partial waves 
to the values closer to the PDG estimates, except for the $N(1900)P_{13}$
resonance, for which the extracted value is still much larger than
the PDG one. Therefore, future investigation should address this problem,
since the $N(1900)P_{13}$ is currently a four-star resonance. 

\begin{table*}[!]
\caption{Fractional decay widths $\sqrt{\Gamma_{\gamma p}\Gamma_{K\Sigma}}/\Gamma_{\mathrm{tot}}$ of
  the nucleon and $\Delta$ resonances 
  extracted from the $\gamma p\to K\Sigma$ channels in the present work and PDG \cite{pdg}.}
\label{tab:brbwproton}
\begin{ruledtabular}
\begin{tabular}{c c c c c c}
Resonances & $J^P$ & \multicolumn{4}{c}{$\sqrt{\Gamma_{\gamma p}\Gamma_{K\Sigma}}/\Gamma_{\mathrm{tot}}$ ($\times 10^{-3}$)}\\
\cline{3-6}
 & & PDG & HFF-P1 & HFF-P3 & HFF-G \\
 \hline
 $N(1700)D_{13}$ & $3/2^-$ &       -       & $0.00  \pm0.00$ & $0.00  \pm0.00$ & $3.84 \pm0.01$ \\
 $N(1710)P_{11}$ & $1/2^+$ & $4.67\pm3.27$ & $1.16  \pm0.07$ & $0.53  \pm0.01$ & $0.00 \pm0.00$ \\
 $N(1720)P_{13}$ & $3/2^+$ &       -       & $81.30 \pm0.02$ & $154.57\pm0.02$ & $13.02\pm0.01$ \\
 $N(1860)F_{15}$ & $5/2^+$ &       -       & $2.68  \pm0.10$ & $2.20  \pm0.08$ & $1.45 \pm0.01$ \\
 $N(1875)D_{13}$ & $3/2^-$ & $0.37\pm0.58$ & $6.46  \pm0.04$ & $6.62  \pm0.03$ & $5.48 \pm0.02$ \\
 $N(1880)P_{11}$ & $1/2^+$ & $3.45\pm5.17$ & $3.48  \pm0.08$ & $2.42  \pm0.04$ & $2.64 \pm0.04$ \\
 $N(1895)S_{11}$ & $1/2^-$ & $2.32\pm3.02$ & $2.94  \pm0.14$ & $6.94  \pm0.04$ & $3.35 \pm0.05$ \\
 $N(1900)P_{13}$ & $3/2^+$ & $4.59\pm7.86$ & $188.71\pm0.06$ & $64.58 \pm0.06$ & $32.75\pm0.04$ \\
 $N(1990)F_{17}$ & $7/2^+$ &       -       & $7.35  \pm0.25$ & $4.56  \pm0.02$ & $2.11 \pm0.02$ \\
 $N(2000)F_{15}$ & $5/2^+$ &       -       & $2.71  \pm0.50$ & $3.20  \pm0.07$ & $3.37 \pm0.02$ \\
 $N(2060)D_{15}$ & $5/2^-$ & $3.93\pm3.90$ & $62.32 \pm0.04$ & $23.88 \pm0.03$ & $6.42 \pm0.03$ \\
 $N(2120)D_{13}$ & $3/2^-$ &       -       & $4.53  \pm0.11$ & $3.44  \pm0.07$ & $3.42 \pm0.08$ \\
 $N(2190)G_{17}$ & $7/2^-$ &       -       & $0.83  \pm0.14$ & $1.22  \pm0.06$ & $0.72 \pm0.04$ \\
 $N(2220)H_{19}$ & $9/2^+$ &       -       & $0.12  \pm0.03$ & $0.07  \pm0.09$ & $0.08 \pm0.07$ \\
 $N(2290)G_{19}$ & $9/2^-$ &       -       & $0.64  \pm0.62$ & $1.30  \pm0.09$ & $0.97 \pm0.09$ \\
 $\Delta(1700)D_{33}$ & $3/2^-$ &       -       & $0.00  \pm0.00$ & $0.00  \pm0.00$ & $1.85 \pm0.01$ \\
 $\Delta(1900)S_{31}$ & $1/2^-$ &       -       & $0.41  \pm0.09$ & $9.35  \pm0.03$ & $11.73\pm0.03$ \\
 $\Delta(1905)F_{35}$ & $5/2^+$ &       -       & $2.03  \pm0.20$ & $0.44  \pm0.15$ & $0.85 \pm0.01$ \\
 $\Delta(1910)P_{31}$ & $1/2^+$ & $1.21\pm1.69$ & $6.16  \pm0.06$ & $0.64  \pm0.04$ & $4.56 \pm0.04$ \\
 $\Delta(1920)P_{33}$ & $3/2^+$ & $4.38\pm5.70$ & $123.09\pm0.07$ & $119.67\pm0.05$ & $48.75\pm0.05$ \\
 $\Delta(1930)D_{35}$ & $5/2^-$ &       -       & $6.57  \pm0.04$ & $18.04 \pm0.04$ & $9.83 \pm0.04$ \\
 $\Delta(1940)D_{33}$ & $3/2^-$ &       -       & $9.95  \pm0.07$ & $4.49  \pm0.04$ & $4.32 \pm0.04$ \\
 $\Delta(1950)F_{37}$ & $7/2^+$ & $0.74\pm1.55$ & $0.99  \pm0.01$ & $6.40  \pm0.06$ & $1.99 \pm0.01$ \\
 $\Delta(2000)F_{35}$ & $5/2^+$ &       -       & $7.62  \pm0.66$ & $2.02  \pm0.04$ & $5.31 \pm0.05$ \\
 $\Delta(2300)H_{39}$ & $9/2^+$ &       -       & $0.44  \pm0.61$ & $0.51  \pm0.14$ & $0.46 \pm0.03$ \\
 $\Delta(2400)G_{39}$ & $9/2^-$ &       -       & $0.71  \pm1.03$ & $1.36  \pm0.18$ & $1.57 \pm0.21$ \\
\end{tabular}
\end{ruledtabular}
\end{table*}

A closer look at Tables~\ref{tab:brbwproton} and \ref{tab:brbwneutron} reveals that
the $N(1720)P_{13}$  resonance experiences the strongest reduction if we use the 
Gaussian form factor (model HFF-G). Compared to the model HFF-P3 the fractional decay
width of this resonance is reduced by more than $90\%$. This is due to the change of 
the resonance role, between the $N(1700)D_{13}$ and $N(1720)P_{13}$ states as we 
mentioned earlier. Since the $N(1720)P_{13}$ is closer to the threshold, the branching 
ratio to this reaction tends to be smaller. In addition to this fact and earlier analysis, 
we have more confidence to say that the $N(1720)P_{13}$ state is more likely to be found 
near the reaction threshold. 


\begin{table*}[!]
\caption{As in Table~\ref{tab:brbwproton}, but extracted from the $\gamma n\to K\Sigma$ channels.}
\label{tab:brbwneutron}
\begin{ruledtabular}
\begin{tabular}{c c c c c c}
Resonances & $J^P$ & \multicolumn{4}{c}{$\sqrt{\Gamma_{\gamma n}\Gamma_{K \Sigma}}/\Gamma_{\mathrm{tot}}$ ($\times 10^{-3}$)}\\
\cline{3-6}
 & & PDG & HFF-P1 & HFF-P3 & HFF-G \\
\hline
 $N(1700)D_{13}$ & $3/2^-$ &       -       & $0.00 \pm0.00$ & $0.00 \pm0.00$ & $2.67 \pm0.01$ \\
 $N(1710)P_{11}$ & $1/2^+$ & $3.74\pm3.74$ & $0.82 \pm0.05$ & $0.13 \pm0.00$ & $0.00 \pm0.00$ \\
 $N(1720)P_{13}$ & $3/2^+$ &       -       & $89.12\pm0.02$ & $62.43\pm0.01$ & $3.40 \pm0.00$ \\
 $N(1860)F_{15}$ & $5/2^+$ &       -       & $3.51 \pm0.13$ & $2.11 \pm0.06$ & $0.49 \pm0.01$ \\
 $N(1875)D_{13}$ & $3/2^-$ & $0.52\pm0.84$ & $6.77 \pm0.02$ & $13.61\pm0.04$ & $9.21 \pm0.02$ \\
 $N(1880)P_{11}$ & $1/2^+$ & $9.84\pm20.1$ & $2.40 \pm0.05$ & $4.35 \pm0.07$ & $2.85 \pm0.05$ \\
 $N(1895)S_{11}$ & $1/2^-$ & $1.88\pm2.62$ & $5.85 \pm0.27$ & $2.96 \pm0.02$ & $1.65 \pm0.03$ \\
 $N(1900)P_{13}$ & $3/2^+$ & $3.87\pm7.73$ & $44.90\pm0.02$ & $25.09\pm0.03$ & $43.97\pm0.06$ \\
 $N(1990)F_{17}$ & $7/2^+$ &       -       & $11.75\pm0.37$ & $6.70 \pm0.03$ & $1.65 \pm0.02$ \\
 $N(2000)F_{15}$ & $5/2^+$ &       -       & $4.28 \pm0.52$ & $4.83 \pm0.11$ & $3.95 \pm0.00$ \\
 $N(2060)D_{15}$ & $5/2^-$ & $2.00\pm2.41$ & $16.23\pm0.01$ & $17.97\pm0.03$ & $4.64 \pm0.04$ \\
 $N(2120)D_{13}$ & $3/2^-$ &       -       & $5.80 \pm0.19$ & $2.18 \pm0.07$ & $6.44 \pm0.14$ \\
 $N(2190)G_{17}$ & $7/2^-$ &       -       & $0.78 \pm0.10$ & $2.61 \pm0.15$ & $0.74 \pm0.03$ \\
 $N(2220)H_{19}$ & $9/2^+$ &       -       & $0.23 \pm0.05$ & $0.13 \pm0.17$ & $0.16 \pm0.13$ \\
 $N(2290)G_{19}$ & $9/2^-$ &       -       & $0.80 \pm0.51$ & $0.37 \pm0.10$ & $1.81 \pm0.17$ \\
\end{tabular}
\end{ruledtabular}
\end{table*}

\section{SUMMARY AND CONCLUSION}
\label{sec:summary}
We have analyzed the $K\Sigma$ photoproduction data for all
four possible isospin channels by using a covariant isobar model
and including all appropriate nucleon and delta resonances.
We used the consistent Lagrangians for hadronic and electromagnetic 
interactions to eliminate the problem of lower-spin background 
contribution. In this analysis, three different form factors have 
been used in the hadronic vertices, i.e., 
the dipole, multi-dipole, and Gaussian ones.
The present model yields a nice agreement between calculated observables 
and experimental data. The best agreement is shown by the models that
employ the multidipole and Gaussian form factors. We have also extracted
the Breit-Wigner masses and widths of the nucleon and delta
resonances as well as 
their masses and widths at their pole positions. By comparing
the extracted values with those given by PDG we conclude that the 
Gaussian form factor leads to a better agreement. This indicates
that the resonances included in the model require a strong 
suppression from the form factors, which is a typical behavior of 
the phenomenological model that employs a large number of resonances.

\section{ACKNOWLEDGMENTS}
The work of S.C. was supported in part by 
the Indonesian Endowment Fund for Education 
(LPDP). T.M. is supported by the PUTI Q2 Grant of Universitas Indonesia, 
under contract No. NKB-1652/UN2.RST/HKP.05.00/2020.

\clearpage

\appendix

\section{The Extracted Masses and Widths}
\label{app:mwbw}
The extracted Breit-Wigner masses and widths of the included nucleon 
and delta resonances in the model are
listed in Table \ref{tab:mwbw}.

\begin{table*}[!]
\centering
\caption{Masses and widths of the nucleon and $\Delta$ resonances 
  obtained from  
  all three models analyzed in the present work.
  \label{tab:mwbw}}
\begin{ruledtabular}
\begin{tabular}{cccccccccc}
 Resonances & $J^P$ & \multicolumn{2}{c}{HFF-P1} & \multicolumn{2}{c}{HFF-P3} & \multicolumn{2}{c}{HFF-G}\\
 \cline{3-4}  \cline{5-6}\cline{7-8}\cline{9-10}
  & & Mass (MeV) & Width (MeV) & Mass (MeV) & Width (MeV) & Mass (MeV) & Width (MeV)\\
 \hline
 $N(1440)P_{11}$ & $1/2^{+}$ & 1420 & 450 & 1450 & 450 & 1450 & 250 \\
 $N(1520)D_{13}$ & $3/2^{-}$ & 1510 & 125 & 1510 & 125 & 1520 & 100 \\
 $N(1535)S_{11}$ & $1/2^{-}$ & 1545 & 125 & 1545 & 125 & 1545 & 175 \\
 $N(1650)S_{11}$ & $1/2^{-}$ & 1670 & 170 & 1661 & 119 & 1648 & 110 \\
 $N(1675)D_{15}$ & $5/2^{-}$ & 1670 & 165 & 1670 & 130 & 1680 & 165 \\
 $N(1680)F_{15}$ & $5/2^{+}$ & 1686 & 120 & 1680 & 140 & 1687 & 120 \\
 $N(1700)D_{13}$ & $3/2^{-}$ & 1650 & 129 & 1650 & 109 & 1750 & 191 \\
 $N(1710)P_{11}$ & $1/2^{+}$ & 1710 & 50  & 1691 & 104 & 1687 & 50  \\
 $N(1720)P_{13}$ & $3/2^{+}$ & 1750 & 400 & 1750 & 248 & 1700 & 361 \\
 $N(1860)F_{15}$ & $5/2^{+}$ & 1829 & 220 & 1820 & 239 & 1820 & 220 \\
 $N(1875)D_{13}$ & $3/2^{-}$ & 1820 & 320 & 1820 & 320 & 1820 & 320 \\
 $N(1880)P_{11}$ & $1/2^{+}$ & 1856 & 180 & 1915 & 216 & 1914 & 180 \\
 $N(1895)S_{11}$ & $1/2^{-}$ & 1893 & 90  & 1893 & 123 & 1893 & 90  \\
 $N(1900)P_{13}$ & $3/2^{+}$ & 1930 & 250 & 1929 & 150 & 1870 & 250 \\
 $N(1990)F_{17}$ & $7/2^{+}$ & 2125 & 400 & 2010 & 200 & 2031 & 200 \\
 $N(2000)F_{15}$ & $5/2^{+}$ & 2044 & 335 & 2030 & 335 & 2030 & 335 \\
 $N(2060)D_{15}$ & $5/2^{-}$ & 2060 & 450 & 2030 & 450 & 2060 & 444 \\
 $N(2120)D_{13}$ & $3/2^{-}$ & 2075 & 305 & 2077 & 305 & 2165 & 305 \\
 $N(2190)G_{17}$ & $7/2^{-}$ & 2200 & 300 & 2200 & 300 & 2105 & 300 \\
 $N(2220)H_{19}$ & $9/2^{+}$ & 2204 & 369 & 2200 & 350 & 2200 & 350 \\
 $N(2290)G_{19}$ & $9/2^{-}$ & 2250 & 300 & 2320 & 300 & 2250 & 300 \\
 $\Delta(1232)P_{33}$ & $3/2^{+}$  & 1230 & 120 & 1234 & 114 & 1234 & 120 \\
 $\Delta(1600)P_{33}$ & $3/2^{+}$  & 1500 & 220 & 1500 & 229 & 1686 & 420 \\
 $\Delta(1620)S_{31}$ & $1/2^{-}$  & 1600 & 150 & 1628 & 150 & 1660 & 150 \\
 $\Delta(1700)D_{33}$ & $3/2^{-}$  & 1686 & 213 & 1686 & 400 & 1750 & 246 \\
 $\Delta(1900)S_{31}$ & $1/2^{-}$  & 1920 & 325 & 1920 & 325 & 1920 & 325 \\
 $\Delta(1905)F_{35}$ & $5/2^{+}$  & 1878 & 400 & 1855 & 270 & 1900 & 270 \\
 $\Delta(1910)P_{31}$ & $1/2^{+}$  & 1910 & 340 & 1910 & 281 & 1910 & 322 \\
 $\Delta(1920)P_{33}$ & $3/2^{+}$  & 1908 & 195 & 1910 & 300 & 1945 & 297 \\
 $\Delta(1930)D_{35}$ & $5/2^{-}$  & 1963 & 220 & 2000 & 223 & 2000 & 220 \\
 $\Delta(1940)D_{33}$ & $3/2^{-}$  & 1994 & 520 & 1954 & 520 & 1940 & 380 \\
 $\Delta(1950)F_{37}$ & $7/2^{+}$  & 1915 & 335 & 1950 & 235 & 1915 & 235 \\
 $\Delta(2000)F_{35}$ & $5/2^{+}$  & 2192 & 525 & 2240 & 275 & 2325 & 525 \\
 $\Delta(2300)H_{39}$ & $9/2^{+}$  & 2393 & 275 & 2429 & 275 & 2405 & 275 \\
 $\Delta(2400)G_{39}$ & $9/2^{-}$  & 2502 & 463 & 2502 & 463 & 2784 & 463 \\
\end{tabular}
\end{ruledtabular}
\end{table*}


\clearpage


\end{document}